\documentclass[draftclsnofoot,onecolumn]{IEEEtran}
\ifCLASSINFOpdf
\else
\fi

\usepackage{amsbsy,amscd,amsfonts,amsgen,amsmath,amsopn,amssymb,amstext,amsthm,amsxtra}
\usepackage{graphicx}
\usepackage{citesort}
\usepackage{color,comment}
\graphicspath{{./} {img/}}
\IEEEoverridecommandlockouts
\newcounter{cnt01}

\begin{document}
\title{Approximate Message Passing Algorithm \\with Universal Denoising \\and Gaussian Mixture Learning}
\author{Yanting~Ma,~\IEEEmembership{Student Member,~IEEE},
Junan~Zhu,~\IEEEmembership{Student Member,~IEEE},
\\and Dror~Baron,~\IEEEmembership{Senior Member,~IEEE}\\
\thanks{This work was supported in part by the National Science Foundation under the Grant CCF-1217749 and in part by the U.S. Army Research Office under the Grant W911NF-14-1-0314. Portions of the work appeared at the 52nd Annual Allerton Conference on Communication, Control, and Computing, Monticello, IL, Oct. 2014~\cite{MaZhuBaronAllerton2014}.}
\thanks{Yanting Ma, Junan Zhu, and Dror Baron are with the Department of Electrical and Computer Engineering, NC State University, Raleigh, NC 27695. E-mail: \{yma7, jzhu9, barondror\}@ncsu.edu.}
}
\maketitle
\thispagestyle{empty}
\begin{abstract}
We study compressed sensing (CS) signal reconstruction
problems where an input signal is measured via 
matrix multiplication under additive white Gaussian noise.
Our signals are assumed to be stationary and ergodic,
but the input statistics are unknown; the goal is to 
provide reconstruction algorithms that are universal
to the input statistics.
We present a novel algorithmic framework that combines:
({\em i}) the approximate message passing (AMP) CS reconstruction framework, which solves the matrix channel recovery problem by iterative scalar channel denoising; 
({\em ii}) a universal denoising scheme based on context quantization, 
which partitions the stationary ergodic signal denoising into
independent and identically distributed (i.i.d.) subsequence denoising; and 
({\em iii}) a density estimation approach that approximates the 
probability distribution of an i.i.d. sequence by fitting a 
Gaussian mixture (GM) model. 
In addition to the algorithmic framework, we provide three contributions: 
({\em i}) numerical results showing that state evolution 
holds for non-separable Bayesian sliding-window denoisers;
({\em ii}) an i.i.d. denoiser based on a modified GM learning algorithm;
and ({\em iii}) a universal denoiser that does not need information about the range where the input takes values from or require the input signal to be bounded.
We provide two implementations of our universal CS recovery algorithm with one being faster and the other being more accurate. The two implementations compare favorably with existing universal reconstruction algorithms in terms of both reconstruction quality and runtime.
\end{abstract}
\IEEEpeerreviewmaketitle

\section{Introduction}
\label{sec:intro}

\subsection{Motivation}
\label{subsec:motivation}
Many scientific and engineering problems can be approximated 
as linear systems of the form
\begin{equation}
{\bf y} = {\bf Ax} + {\bf z},
\label{eq:matrix_channel}
\end{equation}
where ${\bf x}\in\mathbb{R}^N$ is the unknown input signal, ${\bf A}\in\mathbb{R}^{M\times N}$ is the matrix that characterizes the linear system, and ${\bf z\in\mathbb{R}}^M$ is measurement noise. The goal is to estimate ${\bf x}$ from the measurements ${\bf y}$ given ${\bf A}$ and statistical information about ${\bf z}$. When $M\ll N$, the setup is known as compressed sensing (CS); by posing a sparsity or compressibility
requirement on the signal, 
it is indeed possible to accurately recover ${\bf x}$ from the ill-posed linear system~\cite{DonohoCS,CandesRUP}. However, we might need $M>N$ when the signal is dense or the noise is substantial.

One popular scheme to solve the CS recovery problem is LASSO~\cite{Tibshirani1996} (also known as basis pursuit denoising~\cite{DonohoBP}): $\widehat{{\bf x}}=\arg\min_{{\bf x}\in\mathbb{R}^N} \frac{1}{2}\|{\bf y}-{\bf Ax}\|_2^2+\gamma \|{\bf x}\|_1$, where $\|\cdot\|_p$ denotes the $\ell_p$-norm, and $\gamma$ reflects a trade-off between the sparsity $\|{\bf x}\|_1$ and residual $\|{\bf y}-{\bf Ax}\|_2^2$. This approach does not require statistical information about ${\bf x}$ and ${\bf z}$, and can be conveniently solved via standard convex optimization tools or the approximate message passing (AMP) algorithm~\cite{DMM2009}. However, the reconstruction quality is often far from optimal in terms of mean square error (MSE). 

Bayesian CS recovery algorithms based on message passing~\cite{CSBP2010,DMM2010ITW1,RanganGAMP2011ISIT} usually achieve better reconstruction quality, but must know the prior for ${\bf x}$.  
For parametric signals with unknown parameters, one can infer the parameters
and achieve the minimum mean square error (MMSE) in some settings; examples include EM-GM-AMP-MOS~\cite{EMGMTSP}, turboGAMP~\cite{turboGAMP}, and adaptive-GAMP~\cite{Kamilov2014}.

Unfortunately, possible uncertainty about the input statistics
may make it difficult to select a 
model class for empirical Bayes algorithms; a 
mismatched model can yield excess mean square error (EMSE) above the MMSE, and the EMSE can get amplified in linear inverse problems~(\ref{eq:matrix_channel}) compared to that in scalar estimation problems~\cite{MaBaronBeirami2015ISIT}.
Our goal 
is to develop universal schemes that approach the optimal Bayesian performance for stationary ergodic signals despite not knowing the input statistics.
Although others have worked on CS algorithms for independent and identically distributed (i.i.d.) signals with unknown distributions~\cite{EMGMTSP}, we are particularly interested in developing algorithms for signals that may not be well approximated by i.i.d. models, because real-world signals often contain dependencies between different entries. 
For example, we will see in Fig.~\ref{fig:AMP_Chirp} that a chirp sound clip is reconstructed 1--2 dB better with models that can capture such dependencies than i.i.d. models applied to sparse transform coefficients.

While approaches based on Kolmogorov 
complexity~\cite{DonohoKolmogorov,DonohoKolmogorovCS2006,JalaliMaleki2011,JalaliMalekiRichB2014} are theoretically appealing for universal signal recovery, they are not computable in practice~\cite{Cover06,LiVitanyi2008}.
Several algorithms based on Markov chain Monte Carlo (MCMC)~\cite{BaronFinland2011,BaronDuarteAllerton2011,JZ2014SSP,ZhuBaronDuarte2014_SLAM}
leverage the fact that for stationary ergodic signals,
both the per-symbol empirical entropy and Kolmogorov complexity converge
asymptotically almost surely to the entropy rate of the signal~\cite{Cover06}, and aim to minimize the empirical entropy.
The best existing implementation of the MCMC approach~\cite{ZhuBaronDuarte2014_SLAM} often achieves an MSE
that is within 3 dB of the MMSE, which resembles a result
by Donoho for universal denoising~\cite{DonohoKolmogorov}.

In this paper, we confine our attention to 
the system model defined in (\ref{eq:matrix_channel}), 
where the input signal ${\bf x}$ is stationary and ergodic.
We merge concepts from AMP~\cite{DMM2009}, 
Gaussian mixture (GM) learning~\cite{FigueiredoJain2002} for density estimation,
and 
universal denoising
for stationary ergodic signals~\cite{Sivaramakrishnan2008,SW_Context2009}.
We call the resulting universal CS recovery algorithm AMP-UD (AMP with a universal denoiser). 
Two implementations of AMP-UD are provided, and they compare favorably with existing universal approaches in terms of reconstruction quality and runtime.

\begin{figure*}[t!]
\setcounter{cnt01}{1}
\center
\includegraphics[width=160mm]{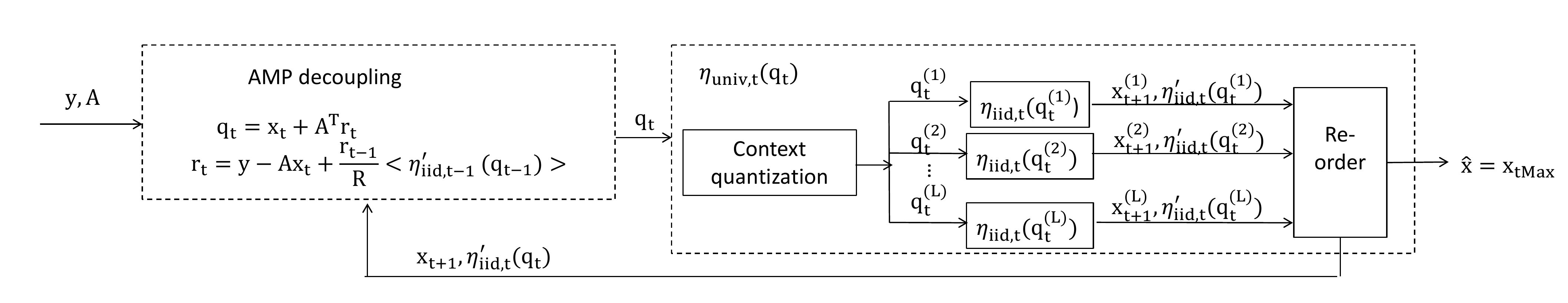}
\caption{Flow chart of AMP-UD. 
AMP~(\ref{eq:AMPiter1}, \ref{eq:AMPiter2}) decouples the linear inverse problem 
into scalar channel denoising problems. In the $t$-th iteration, the universal denoiser 
$\eta_{\text{univ},t}(\cdot)$ converts stationary ergodic signal denoising into i.i.d. subsequence denoising. Each i.i.d. denoiser $\eta_{\text{iid},t}(\cdot)$~(\ref{eq:FJ_denoiser}) outputs the denoised subsequence ${\bf x}_{t+1}^{(l)}$ and the derivative of the denoiser $\eta_{\text{iid},t}'(\cdot)$~(\ref{eq:denoiser_FJ_deriv}). The algorithm stops when the iteration index $t$ reaches the predefined maximum tMax, and outputs 
$\widehat{\bf x}_{\text{tMax}}$ as the CS recovery result.}
\label{fig:FlowChart}
\setcounter{figure}{\value{cnt01}}
\end{figure*}

\subsection{Related work and main results}
\label{subsec:contribution}
{\bf Approximate message passing:} 
AMP is an iterative algorithm that solves a linear inverse problem by successively converting matrix channel problems into scalar channel denoising
problems with additive white Gaussian noise (AWGN). AMP has received considerable attention, because of its fast convergence and the state evolution (SE) formalism~\cite{DMM2009,Bayati2011}, which offers a precise
characterization of the AWGN denoising problem in each iteration. 
AMP with separable denoisers has been rigorously proved to obey SE~\cite{Bayati2011}. 

The focus of this paper is the reconstruction of signals that are not necessarily i.i.d., and so we need to explore non-separable denoisers.
Donoho {\em et al.}~\cite{Donoho2013} provide numerical results demonstrating 
that SE accurately predicts the phase transition of AMP when some well-behaved non-separable minimax denoisers are applied, and conjecture that SE holds for AMP with a broader class of denoisers. 
A compressive imaging algorithm that applies non-separable image denoisers within AMP appears in Tan {\em et al.}~\cite{Tan_CompressiveImage2014}.
Rush {\em et al.}~\cite{Rush_ISIT2015} 
apply AMP to sparse superposition decoding, and prove that SE holds for AMP with certain block-separable denoisers and that such an AMP-based decoder achieves channel capacity.
A potential challenge of implementing AMP is to obtain the
Onsager correction term~\cite{DMM2009}, 
which involves the calculation of the derivative of a denoiser.
Metzler {\em et al.}~\cite{Metzler2014} leverage a Monte Carlo technique to approximate the derivative of a denoiser when an explicit analytical formulation of the denoiser is unavailable, and provide numerical results showing that SE holds for AMP with their approximation.

Despite the encouraging results for using non-separable denoisers within AMP,
a rigorous proof that SE holds for general non-separable denoisers has yet to appear.
Consequently, new evidence showing that AMP obeys SE 
may increase the community's confidence about using non-separable 
denoisers within AMP. 
{\em Our first contribution is that we provide numerical results showing that SE holds for non-separable Bayesian sliding-window denoisers. }

{\bf Fitting Gaussian mixture models:}
Figueiredo and Jain~\cite{FigueiredoJain2002} propose an 
unsupervised GM learning
algorithm that fits 
a given data sequence with a GM model. The algorithm employs a cost function
that resembles the minimum message length 
criterion, and the parameters
are learned 
via expectation-maximization (EM).

Our GM fitting problem involves estimating the  
probability density function (pdf) of a sequence ${\bf x}$
from its AWGN corrupted observations.
We modify the GM fitting algorithm~\cite{FigueiredoJain2002}, so that a GM model can be
learned from noisy data. 
Once the estimated pdf $\widehat{p}_X$ of ${\bf x}$ is available, we 
estimate ${\bf x}$ by computing 
the conditional expectation
with the estimated pdf $\widehat{p}_X$ (recall that MMSE estimators rely on conditional expectation).
{\em Our second contribution is that we modify the GM learning algorithm, and extend it to an i.i.d. denoiser.}

{\bf Universal denoising:} 
Our denoiser for stationary ergodic signals is inspired by a context 
quantization approach~\cite{SW_Context2009}, where a universal denoiser for a 
stationary ergodic signal involves multiple i.i.d. denoisers for 
conditionally i.i.d. subsequences. Sivaramakrishnan and 
Weissman~\cite{SW_Context2009} have shown 
that their universal denoiser based on context quantization 
can achieve the MMSE asymptotically for stationary ergodic signals with known bounds.

The boundedness condition of Sivaramakrishnan and 
Weissman~\cite{SW_Context2009}
is partly due to their density estimation approach, 
in which the empirical distribution function
is obtained by quantizing the bounded range of the signal.
Such boundedness conditions may be undesirable in certain applications. 
We overcome this limitation by replacing their density estimation approach 
with GM model learning.  
{\em Our third contribution is a universal denoiser that does not need information about the bounds or require the input signal to be bounded; we conjecture that our universal denoiser achieves the MMSE asymptotically under some technical conditions.}

A flow chart of AMP-UD, which employs the AMP framework, along with our modified universal denoiser ($\eta_{\text{univ}}$) and the GM-based i.i.d. denoiser ($\eta_{\text{iid}}$), is shown in Fig.~\ref{fig:FlowChart}. 
Based on the numerical evidence that SE holds for AMP with Bayesian sliding-window denoisers and the conjecture that our universal denoiser can achieve the MMSE, we further conjecture that AMP-UD achieves the MMSE under some technical conditions.
The details of AMP-UD, including two practical implementations, are developed in Sections~\ref{sec:AMP}--\ref{sec:proposed_algo}.

The remainder of the paper is arranged as follows. 
In Section~\ref{sec:AMP}, we review AMP and provide new numerical evidence 
that AMP obeys SE with non-separable denoisers. 
Section~\ref{sec:FJ} modifies the GM fitting algorithm, and extends it to an i.i.d. denoiser.
In Section~\ref{sec:SW}, we extend the universal denoiser based on
context quantization to overcome the boundedness condition, and two implementations are provided to improve denoising quality. 
Our proposed AMP-UD algorithm is summarized in Section~\ref{sec:proposed_algo}. 
Numerical results are shown in Section~\ref{sec:numerical}, and we conclude the paper in Section~\ref{sec:conclusion}.

\section{Approximate message passing with sliding-window denoisers}
In this section, we apply non-separable Bayesian sliding-window denoisers within AMP, and provide numerical evidence that state evolution (SE) holds for AMP with this class of denoisers.
\label{sec:AMP}

\subsection{Review of AMP}
\label{subsec:AMP_review}
Consider a linear 
system~(\ref{eq:matrix_channel}), where the measurement matrix~${\bf A}$ has zero mean i.i.d. Gaussian entries with unit-norm columns on average,
and ${\bf z}$ represents i.i.d. Gaussian noise with pdf $p_Z(z_i)=\mathcal{N}(z_i;0,\sigma_z^2)$, where $z_i$ is the $i$-th entry of the vector ${\bf z}$, and $\mathcal{N}(x;\mu,\sigma^2)$ denotes a Gaussian pdf:
\begin{equation*}
\mathcal{N}(x;\mu,\sigma^2)=\frac{1}{\sqrt{2\pi}\sigma}\exp\left(-\frac{(x-\mu)^2}{2\sigma^2}\right).
\end{equation*}
Note that 
AMP has been proved to follow SE when ${\bf A}$ is a zero mean i.i.d. Gaussian matrix, but may diverge otherwise. Several techniques have been proposed to improve the convergence of AMP~\cite{Rangan2014ISIT,Swamp2014,Vila2015,RanganADMMGAMP2015_ISIT}.
Moreover, other noise distributions can be supported using generalized AMP (GAMP)~\cite{RanganGAMP2011ISIT}, and the noise distribution can be estimated in each GAMP iteration~\cite{Kamilov2014}. Such generalizations 
are beyond the scope of this work. 

Starting with ${\bf x}_0={\bf 0}$, the AMP algorithm~\cite{DMM2009} proceeds iteratively according to
\begin{align}
{\bf x}_{t+1}&=\eta_t({\bf A}^T{\bf r}_t+{\bf x}_t)\label{eq:AMPiter1},\\
{\bf r}_t&={\bf y}-{\bf Ax}_t+\frac{1}{R}{\bf r}_{t-1}
\langle\eta_{t-1}'({\bf A}^T{\bf r}_{t-1}+{\bf x}_{t-1})\rangle\label{eq:AMPiter2},
\end{align}
where~$R=M/N$ represents the measurement rate, $t$ represents the iteration index, $\eta_t(\cdot)$ is a denoising function, and~$\langle{\bf u}\rangle=\frac{1}{N}\sum_{i=1}^N u_i$
for some vector~${\bf u}\in\mathbb{R}^N$. 
The last term in (\ref{eq:AMPiter2}) is called the Onsager correction term in statistical physics.
The empirical distribution of ${\bf x}$ is assumed to converge to some probability distribution $p_X$ on $\mathbb{R}$, and
the denoising function~$\eta_t(\cdot)$ is separable in the original derivation of AMP~\cite{DMM2009,Montanari2012,Bayati2011}. That is, $\eta_t({\bf u})=(\eta_t(u_1),\eta_t(u_2),...,\eta_t(u_N))$ and $\eta_{t}'({\bf u})=(\eta_{t}'(u_1),\eta_{t}'(u_2),...,\eta_{t}'(u_N))$, where $\eta_{t}'(\cdot)$ denotes the derivative of $\eta_t(\cdot)$. 
A useful property of AMP is that at each iteration, 
the vector~${\bf A}^T{\bf r}_t+{\bf x}_t\in\mathbb{R}^N$ 
in (\ref{eq:AMPiter1}) is statistically equivalent to 
the input signal ${\bf x}$ corrupted by AWGN, where the noise variance~$\sigma_t^2$ evolves following SE in the limit of large systems ($N\rightarrow\infty, M/N\rightarrow R$):
\begin{equation}
\sigma_{t+1}^2=\sigma^2_z+\frac{1}{R}\text{MSE}(\eta_t,\sigma_t^2)\label{eq:SE},
\end{equation}
where $\text{MSE}(\eta_t,\sigma_t^2)=\mathbb{E}_{X,W}\left[\left( \eta_t\left( X+\sigma_{t}W \right)-X \right)^2\right]$, $W\sim\mathcal{N}(w;0,1)$, $X\sim p_X$, and $\sigma_0^2=\sigma_z^2+\frac{1}{R}\mathbb{E}[X^2]$.
Formal statements for SE appear 
in the reference papers~\cite{Bayati2011,Montanari2012}. 
Additionally, it is convenient to use the following
estimator for $\sigma_t^2$~\cite{Bayati2011,Montanari2012}:
\begin{equation}
\widehat{\sigma}_t^2 = \frac{1}{M}\|{\bf r}_t\|_2^2.
\label{eq:AMP_noise_est}
\end{equation}

\subsection{State evolution for Bayesian sliding-window denoisers}
\label{subsec:SE}
SE allows to calculate the asymptotic MSE of linear systems from the MSE of the denoiser used within AMP. Therefore, knowing that SE holds for AMP with the denoisers that we are interested in can help us choose a good denoiser for AMP.
It has been conjectured by Donoho {\em et al.}~\cite{Donoho2013} that 
AMP with a wide range of non-separable denoisers obeys SE.
We now provide new evidence to support this conjecture by constructing 
non-separable Bayesian denoisers within a sliding-window denoising scheme for two stationary ergodic Markov signal models, and showing that SE accurately predicts the performance of AMP with this class of denoisers for large signal dimension $N$. Note that for a signal that is generated by a stationary ergodic process, its empirical distribution converges to the stationary distribution, hence the condition on the input signal in the proof for SE~\cite{Bayati2011} is satisfied, and our goal is to numerically verify that SE holds for AMP with non-separable sliding-window denoisers for stationary ergodic signal models.
Our rationale for examining the SE performance of sliding-window
denoisers is that the context quantization based universal denoiser
~\cite{SW_Context2009}, which will be used in Section~\ref{sec:SW}, resembles
a sliding-window denoiser. 
 
The mathematical model for an AWGN channel denoising problem is defined as
\begin{equation}
{\bf q}={\bf x}+{\bf v},
\label{eq:denoising}
\end{equation}
where ${\bf x}\in\mathbb{R}^N$ is the input signal, ${\bf v}\in\mathbb{R}^N$ is AWGN with pdf $p_V(v_i)=\mathcal{N}(v_i;0,\sigma_v^2)$, and ${\bf q}\in\mathbb{R}^N$ is a sequence of noisy observations. 
Note that we are interested in designing denoisers for AMP,
and the noise variance of the scalar channel in each AMP iteration can be estimated as $\widehat{\sigma}_t^2$~(\ref{eq:AMP_noise_est}). Therefore, throughout the paper
we assume that the noise variance $\sigma_v^2$ is known when we discuss scalar channels. 

In a separable
denoiser, $x_j$ is estimated only from its noisy observation $q_j$. The separable Bayesian denoiser that minimizes the MSE is point-wise conditional expectation,
\begin{equation}
\widehat{x}_j=\mathbb{E}[X|Q=q_j]=\int xp(x|q_j)\text{d}x,\label{eq:Bayes_scalar}
\end{equation}
where Bayes' rule yields $p(x|q_j)=\frac{\mathcal{N}(q_j;x,\sigma_v^2)p_X(x)}{p_Q(q_j)}$. If entries of the input signal ${\bf x}$ are drawn independently from $p_X$, then (\ref{eq:Bayes_scalar}) achieves the MMSE.

When there are statistical dependencies among the entries of ${\bf x}$, a sliding-window 
scheme can be applied to improve the MSE.
We consider two Markov sources as examples that contain statistical
dependencies, and emphasize that our true motivation is the richer 
class of stationary ergodic sources.

{\bf Example source~1:} Consider a two-state Markov state machine that
contains states $s_0$ (zero state in which the signal entries are zero) and $s_1$ (nonzero state in which entries are nonzero). 
The transition probabilities are $p_{10}=p(s_0|s_1)$ and 
$p_{01}=p(s_1|s_0)$. In the steady state, the marginal 
probability of state $s_1$ is $\frac{p_{01}}{p_{01}+p_{10}}$.
We call our first example source Markov-Gaussian (MGauss for short);
it is generated by the two-state Markov machine with $p_{01}=\frac{3}{970}$ 
and $p_{10}=\frac{1}{10}$, 
and in the nonzero state the signal value follows a Gaussian distribution $\mathcal{N}(x;\mu_x,\sigma_x^2)$.
These state transition parameters yield 3\% nonzero 
entries in an MGauss signal on average. 

{\bf Example source~2:} Our second example is
a four-state Markov switching signal (M4 for short) that follows 
the pattern $+1,+1,-1,-1,+1,+1,-1,-1...$ with 3\% error probability in state transitions, resulting in the signal switching from $-1$ to $+1$ or vice versa either too early or too late; the four states $s_1=[-1\ -1]$, $s_2=[-1\ +1]$, $s_3=[+1\ -1]$, 
and $s_4=[+1\ +1]$ have equal marginal probabilities $0.25$ in 
the steady state.

{\bf Bayesian sliding-window denoiser:}
Let $\boldsymbol\theta$ be a binary vector, where $\theta_i=0$ indicates $x_i=0$, and $\theta_i=1$ indicates $x_i \neq 0$. Denoting a block $(u_s, u_{s+1}, ...,u_t)$ of any sequence ${\bf u}$ by ${\bf u}_s^t$ for $s < t$, the $(2k + 1)$-Bayesian sliding-window denoiser $\eta_\text{MGauss}$ for the MGauss signal is defined as
\begin{align}
\eta_{\text{MGauss},j}({\bf q}_{j-k}^{j+k})&=\mathbb{E}[X_j|{\bf Q}_{j-k}^{j+k}={\bf q}_{j-k}^{j+k}]\nonumber\\
&=\frac{\sum\limits_{\substack{{\boldsymbol \theta}_{j-k}^{j+k} \in\{s_0,s_1\}^{2k+1}\\ \theta_j=s_1}}
\left(\prod\limits_{i=j-k}^{j+k}h(q_i,\theta_i;\mu_x,\sigma_x^2,\sigma_v^2) 
p_{{\boldsymbol \Theta}_{j-k}^{j+k}}({\boldsymbol\theta}_{j-k}^{j+k})\right)}
{p_{{\bf Q}_{j-k}^{j+k}}({\bf q}_{j-k}^{j+k})}\cdot\left(\frac{\sigma_x^2}{\sigma_x^2+\sigma_v^2}(q_j-\mu_x)+\mu_x\right),\label{eq:Bayes_MGauss}
\end{align}
where 
\begin{align*}
h(q_i,\theta_i;\mu_x,\sigma_x^2,\sigma_v^2)
&=\begin{cases}
\mathcal{N}(q_i;\mu_x,\sigma_v^2+\sigma_x^2),\quad &\text{if }\theta_i=s_1\\
\mathcal{N}(q_i;0,\sigma_v^2),\quad &\text{if } \theta_i=s_0
\end{cases},\\
p_{{\boldsymbol\Theta}_{j-k}^{j+k}}({\boldsymbol\theta}_{j-k}^{j+k})&=p(\theta_{j-k})\prod\limits_{i=j-k}^{j+k-1}p(\theta_{i+1}|\theta_{i}),\\
p_{{\bf Q}_{j-k}^{j+k}}({\bf q}_{j-k}^{j+k})
&=\sum p_{{\boldsymbol \Theta}_{j-k}^{j+k}}({\boldsymbol\theta}_{j-k}^{j+k})\prod\limits_{i=j-k}^{j+k}h(q_i,\theta_i;\mu_x,\sigma_x^2),
\end{align*}
and the summation is over ${\boldsymbol \theta}_{j-k}^{j+k}\in\{s_0,s_1\}^{2k+1}$.

The MSE of $\eta_{\text{MGauss},j}$ is
\begin{align}
\text{MSE}(\eta_\text{MGauss},\sigma_v^2)&=\mathbb{E}\left[\left(X_j-\eta_{\text{MGauss},j}({\bf Q}_{j-k}^{j+k})\right)^2\right]\nonumber\\
&=\frac{p_{01}(\sigma_x^2+\mu_x^2)}{p_{01}+p_{10}}-\int_{\mathbb{R}^{2k+1}} \eta_{\text{MGauss},j}^2({\bf q})p_{{\bf Q}_{j-k}^{j+k}}({\bf q})d {\bf q}.\label{eq:Bayes_MGauss_MSE}
\end{align}

Similarly, the $(2k+1)$-Bayesian sliding-window denoiser $\eta_\text{M4}$ for the M4 signal is defined as
\begin{align}
\eta_{\text{M4},j}({\bf q}_{j-k}^{j+k})&=\mathbb{E}[X_j|{\bf Q}_{j-k}^{j+k}={\bf q}_{j-k}^{j+k}]\nonumber\\
&=\frac{p_{X_j,{\bf Q}_{j-k}^{j+k}}(1,{\bf q}_{j-k}^{j+k})-p_{X_j,{\bf Q}_{j-k}^{j+k}}(-1,{\bf q}_{j-k}^{j+k})}
{p_{X_j,{\bf Q}_{j-k}^{j+k}}(1,{\bf q}_{j-k}^{j+k})+p_{X_j,{\bf Q}_{j-k}^{j+k}}(-1,{\bf q}_{j-k}^{j+k})},\label{eq:Bayes_M4}
\end{align}
where 
\begin{align*}
p_{{\bf X}_{j-k}^{j+k}}({\bf x}_{j-k}^{j+k})&=p(x_{j-k},x_{j-k+1})\prod_{i=j-k}^{j+k-2}p(x_{i+2}|x_{i+1},x_{i}),\\
p_{X_j,{\bf Q}_{j-k}^{j+k}}(x,{\bf q}_{j-k}^{j+k})&=\sum p_{{\bf X}_{j-k}^{j+k}}({\bf x}_{j-k}^{j+k}) \prod\limits_{i=j-k}^{j+k}\mathcal{N}(q_i;x_i,\sigma_v^2),
\end{align*}
where the summation is over ${\bf x}_{j-k}^{j+k}\in\{-1,1\}^{2k+1}$ with $x_j=x\in\{-1,1\}$ fixed.

It can be shown that
\begin{align}
\text{MSE}(\eta_{\text{M4}},\sigma_v^2)&=\mathbb{E}\left[\left(X_j-\eta_{\text{M4},j}({\bf Q}_{j-k}^{j+k})\right)^2\right]\nonumber\\
&= 4\int_{\mathbb{R}^{2k+1}}\frac{p_{X_j,{\bf Q}_{j-k}^{j+k}}(-1,{\bf q})p_{X_j,{\bf Q}_{j-k}^{j+k}}(1,{\bf q})}{p_{{\bf Q}_{j-k}^{j+k}}({\bf q})} \text{d}{\bf q}.\label{eq:Bayes_M4_MSE}
\end{align}

If AMP with $\eta_\text{MGauss}$ or $\eta_\text{M4}$ obeys SE, 
then the noise variance $\sigma_t^2$ should evolve 
according to (\ref{eq:SE}). As a consequence, 
the reconstruction error at iteration $t$ can be predicted by evaluating (\ref{eq:Bayes_MGauss_MSE}) or (\ref{eq:Bayes_M4_MSE}) with $\sigma_v^2$ being replaced by $\sigma_t^2$.

\begin{figure}[t!]
\center
\includegraphics[width=85mm]{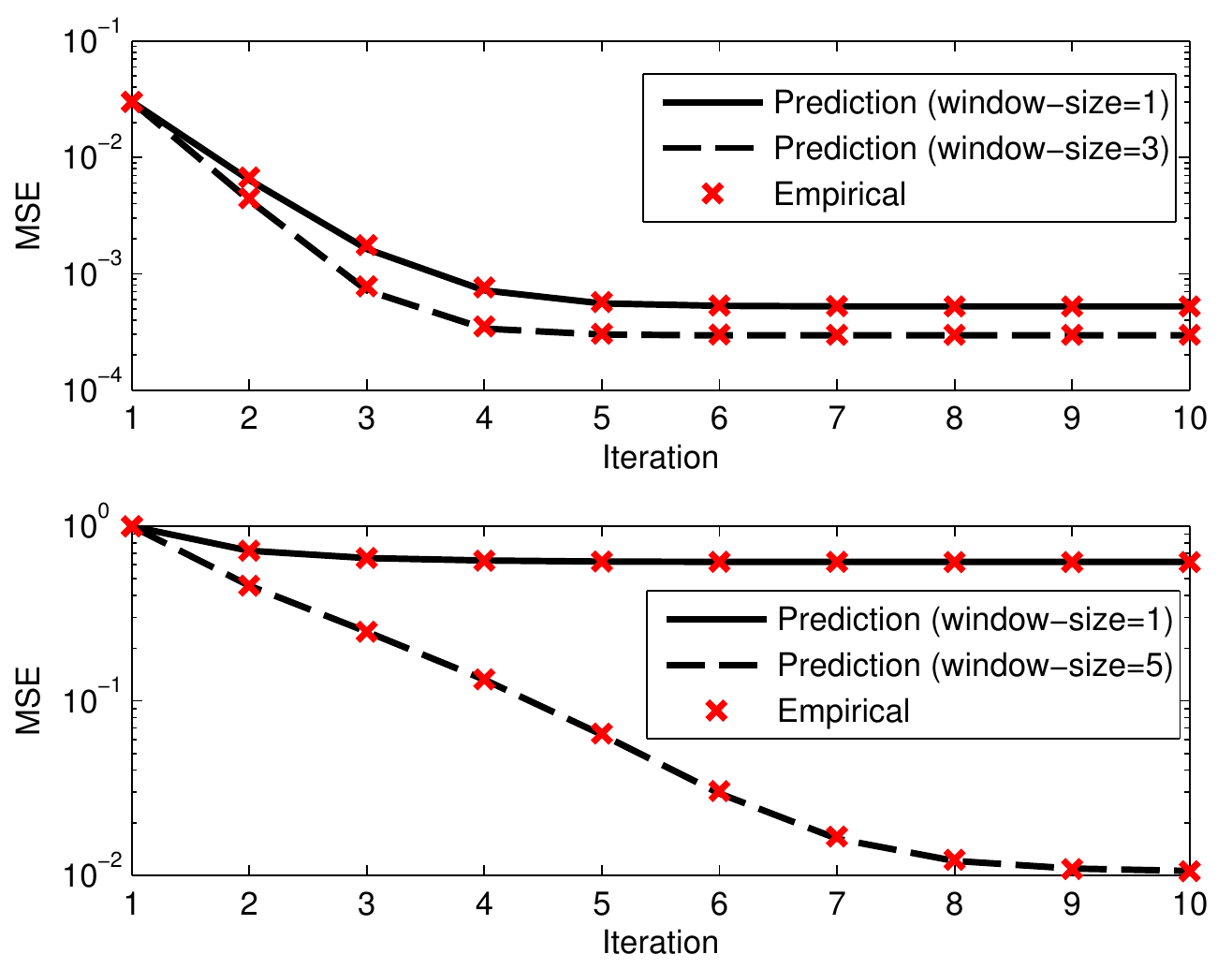}
\caption{Top: Numerical verification of SE for AMP with $\eta_{\text{MGauss}}$ (\ref{eq:Bayes_MGauss}) when the input is an MGauss signal. ($N=20000, R=M/N=0.4, \text{SNR}=10 \text{ dB}$.) Bottom: Numerical verification of SE for AMP with $\eta_{\text{M4}}$ (\ref{eq:Bayes_M4}) when the input is an M4 signal. ($N=20000, R=0.4, \text{SNR}=10 \text{ dB}$.)}
\label{fig:SE_verify}
\end{figure}

{\bf Numerical evidence:} We apply $\eta_{\text{MGauss}}$ (\ref{eq:Bayes_MGauss}) within AMP 
for MGauss signals, and $\eta_{\text{M4}}$ (\ref{eq:Bayes_M4}) within
AMP for M4 signals. The window size $2k+1$ is chosen to be 1 or 3 for $\eta_{\text{MGauss}}$, and 1 or 5 for  $\eta_{\text{M4}}$. Note that when the window size is 1, $\eta_{\text{MGauss}}$ and  $\eta_{\text{M4}}$ become separable denoisers. The MSE predicted by SE is compared to the empirical MSE at each iteration where the input signal to noise ratio ($\text{SNR}=10\log_{10}[(N\mathbb{E}[X^2])/(M\sigma_z^2)]$) is 
10 dB for both MGauss and M4.
It is shown in Fig.~\ref{fig:SE_verify} 
for AMP with $\eta_{\text{MGauss}}$ and $\eta_{\text{M4}}$
that the markers representing the empirical MSE track the lines predicted by SE, 
and that side-information from neighboring entries helps 
improve the MSE.

Our SE results for the two Markov sources increase
our confidence that
AMP with non-separable denoisers that incorporate information from neighboring entries will track SE.

The reader may have noticed from Fig.~\ref{fig:FlowChart} that the universal denoiser $\eta_\text{univ}(\cdot)$ is acting as a set of separable denoisers $\eta_\text{iid}(\cdot)$. 
However, the statistical information used by $\eta_\text{iid}(\cdot)$ is learned from subsequences ${\bf q}_t^{(1)}$,...,${\bf q}_t^{(L)}$ of the noisy sequence ${\bf q}_t$, and the subsequencing result
is determined by the neighborhood of each entry.
The SE results for the Bayesian sliding-window denoisers motivate us to apply the universal denoiser
within AMP for CS reconstruction of stationary ergodic signals with
unknown input statistics. 
Indeed, the numerical results in Section~\ref{sec:numerical} show that AMP with a universal denoiser leads to a promising universal CS recovery algorithm.

\section{i.i.d. denoising via Gaussian mixture fitting}
\label{sec:FJ}
We will see in Section~\ref{sec:SW} that context quantization maps the non-i.i.d. sequence ${\bf q}$ into conditionally independent subsequences, and now focus our attention on denoising the resulting i.i.d. subsequences.

\subsection{Background}
\label{subsec:FJ_old}
The pdf of a Gaussian mixture (GM) has the form:
\begin{equation}
p(x) = \sum_{s=1}^{S}\alpha_s\mathcal{N}(x;\mu_s,\sigma_s^2),\label{eq:GM}
\end{equation}
where $S$ is the number of Gaussian components, and $\sum_{s=1}^S\alpha_s=1$, so that $p(x)$ is a proper pdf. 

Figueiredo and Jain~\cite{FigueiredoJain2002} propose to fit a GM model for a given data sequence by starting with some arbitrarily large $S$, and inferring the structure of the mixture by letting the mixing probabilities $\alpha_s$ of some components be zero. This leads to an unsupervised learning algorithm that automatically determines the number of Gaussian components from data. This approach resembles the concept underlying the minimum message length (MML) criterion that selects the best overall model from the entire model space, which differs from
model class selection based on the best model within each class.\footnote{All models with the same number of components belong to one model class, and different models within a model class have different parameters for each component.} 
This criterion can be interpreted as posing a Dirichlet prior on the mixing probability and perform maximum a poteriori estimation~\cite{FigueiredoJain2002}.
A component-wise EM algorithm that updates $\{\alpha_s,\mu_s,\sigma_s^2\}$ sequentially in $s$ is used to implement the MML-based approach.
The main feature of the component-wise EM algorithm is that if $\alpha_s$ is estimated as 0, then the $s$-th component is immediately removed, and the expectation is recalculated before moving to the estimation of the next component.
\subsection{Extension to denoising}
\label{subsec:FJ_new}
Consider the scalar channel denoising problem defined in (\ref{eq:denoising})
with an i.i.d. input. We propose to estimate ${\bf x}$ from its Gaussian noise corrupted observations ${\bf q}$ by posing a GM prior on ${\bf x}$, and learning the parameters of the GM model with a modified version of the algorithm  
by Figueiredo and Jain~\cite{FigueiredoJain2002}.

{\bf Initialization of EM:}
The EM algorithm must be initialized for each parameter, $\lbrace \alpha_s,\mu_s,\sigma_s^2 \rbrace$, $s=1,...,S$. 
One may choose to initialize the Gaussian components with equal mixing probabilities and equal variances, and the initial value of the means are 
randomly sampled from the input data sequence~\cite{FigueiredoJain2002}, which in our case is the sequence of noisy observations ${\bf q}$.
However, in CS recovery problems, the input signal is often sparse, 
and it becomes difficult to correct the initial value if the initialized
values are far from the truth. 
To see why a poor initialization might be problematic,
consider the following scenario: a sparse binary signal that contains a few ones and is corrupted by Gaussian noise is sent to the algorithm. If the initialization levels of the {\em $\mu_s$}'s are all around zero, then the algorithm is likely to fit a Gaussian component with near-zero mean and large variance rather than two narrow Gaussian components, one of which has mean close to zero while the other has mean close to one. 

To address this issue, we modify the initialization to examine the maximal distance between each symbol of the input data sequence and the current initialization of the {\em $\mu_s$}'s. If the distance is greater than $0.1\sigma_{q}$, then we add a Gaussian component whose mean is initialized as the value of the symbol being examined, where 
$\sigma_{q}^2$ is the estimated variance of the noisy observations ${\bf q}$. We found in our simulations that the modified initialization improves the accuracy of the density estimation, and speeds up the convergence of the EM algorithm; the details of the simulation are omitted for brevity. 

{\bf Parameter estimation from noisy data:}
Two possible modifications can be made to the original GM learning algorithm~\cite{FigueiredoJain2002} that is designed for clean data.

We first notice that the model for the noisy data is a GM convolved with Gaussian noise, which is a new GM with larger component variances. Hence, one approach is to use the original algorithm~\cite{FigueiredoJain2002} to fit a GM to the noisy data, but to remove a component immediately during the EM iterations if the estimated component variance is much smaller than the noise variance $\sigma_v^2$. Specifically, during the parameter learning process, if a component has variance that is
less than $0.2\sigma_v^2$, we assume that this
low-variance component is spurious, and remove it from the mixture model. 
However, if the component variance is between $0.2\sigma_v^2$ 
and $0.9\sigma_v^2$, then we force the component variance to be $0.9\sigma_v^2$ and let the algorithm keep tracking this component. For component variance greater than $0.9\sigma_v^2$, we do not adjust the algorithm. 
The parameters 0.2 and 0.9 are chosen, because they provide reasonable MSE performance for a wide range of signals that we tested. These parameters are then fixed for our algorithm to generate the numerical results in Section~\ref{sec:numerical}.
At the end of the parameter learning process, all remaining components with variances less than $\sigma_v^2$ are set to have variances equal to $\sigma_v^2$.
That said, when subtracting the noise variance $\sigma_v^2$ from the Gaussian components of $\widehat{p}_Q$ to obtain the components of 
$\widehat{p}_X$, we could have components with zero-valued variance, which yields deltas in $\widehat{p}_X$. 
Note that deltas are in general difficult to fit with a limited amount of observations, and our modification helps the algorithm estimate deltas.

Another approach is to introduce latent variables that represent the underlying clean data, and estimate the parameters of the GM for the latent variables directly. Hence, similar to the original algorithm, a component is removed only when the estimated mixing probability is non-positive. It can be shown that the GM parameters are estimated as
\begin{align*}
\widehat{\alpha}_s(t+1) &= \frac{\max\Big\lbrace\sum\limits_{i=1}^N w_i^{(s)}(t)-1,0\Big\rbrace}{\sum\limits_{s:\alpha_s>0}\max\Big\lbrace\sum\limits_{i=1}^N w_i^{(s)}(t)-1,0\Big\rbrace},\\
\widehat{\mu}_s(t+1)&=\frac{\sum\limits_{i=1}^N w_i^{(s)}(t)a_i^{(s)}(t)}{\sum\limits_{i=1}^Nw_i^{(s)}(t)},\\
\widehat{\sigma}_s^2(t+1)&=\frac{\sum\limits_{i=1}^N w_i^{(s)}(t)\left(v_i^{(s)}(t)+\left(a_i^{(s)}(t)-\widehat{\mu}_s(t+1)\right)^2\right)}{\sum\limits_{i=1}^Nw_i^{(s)}(t)},
\end{align*}
where
\begin{align*}
w_i^{(s)}(t)&=\frac{\widehat{\alpha}_s(t)\mathcal{N}(q_i;\widehat{\mu}_s(t),\sigma_v^2+\widehat{\sigma}_s(t)^2)}{\sum\limits_{m=1}^S\widehat{\alpha}_m(t)\mathcal{N}(q_i;\widehat{\mu}_m(t),\sigma_v^2+\widehat{\sigma}_m^2(t))},\\
a_i^{(s)}(t)&=\frac{\widehat{\sigma}_s^2(t)}{\widehat{\sigma}_s^2(t)+\sigma_v^2}(q_i-\widehat{\mu}_s(t))+\widehat{\mu}_s(t),\\
v_i^{(s)}(t)&=\frac{\sigma_v^2\sigma_s^2}{\sigma_v^2+\widehat{\sigma}_s^2(t)}.
\end{align*}
Detailed derivations appear in the Appendix.

We found in our simulation that the first approach converges faster and leads to lower reconstruction error, especially for discrete-valued inputs. Therefore, the simulation results presented in Section~\ref{sec:numerical} use the first approach.

{\bf Denoising:} 
Once the parameters in (\ref{eq:GM}) are estimated, 
we define a denoiser for i.i.d. signals as conditional expectation:
\begin{align}
&\eta_{\text{iid}}(q) =\mathbb{E}[X|Q=q]\nonumber\\
&=\sum_{s=1}^S\mathbb{E}[X|Q=q,\text{comp}=s]P(\text{comp}=s|Q=q)\nonumber\\
&=\sum_{s=1}^S\left(\frac{\sigma_s^2}{\sigma_s^2+\sigma_v^2}(q-\mu_s)+\mu_s\right)\frac{\alpha_s\mathcal{N}(q;\mu_s,\sigma_s^2+\sigma_v^2)}{\sum_{s=1}^S\alpha_s\mathcal{N}(q;\mu_s,\sigma_s^2+\sigma_v^2)},\label{eq:FJ_denoiser}
\end{align}
where comp is the component index, and 
\begin{equation*}
\mathbb{E}[X|Q=q,\text{comp}=s]=\left(\frac{\sigma_s^2}{\sigma_s^2+\sigma_v^2}(q-\mu_s)+\mu_s\right)
\end{equation*}
is the Wiener filter for component $s$.

We have verified numerically for several distributions and low to moderate noise levels that the denoising results obtained by the GM-based i.i.d. denoiser~(\ref{eq:FJ_denoiser}) approach the 
MMSE within a few hundredths of a dB. For example, the favorable reconstruction results for i.i.d. sparse Laplace signals in Fig.~\ref{fig:AMP_Laplace} show that the GM-based denoiser approaches the MMSE.

\section{Universal denoising}
\label{sec:SW}
We have seen in Section~\ref{sec:FJ} that an i.i.d. denoiser based on GM learning can denoise i.i.d. signals with unknown distributions. 
Our goal in this work is to reconstruct stationary ergodic signals that are not necessarily i.i.d.
Sivaramakrishnan and Weissman~\cite{SW_Context2009} have proposed a universal denoising scheme for stationary ergodic signals with known bounds based on context quantization, where a stationary ergodic signal is partitioned into 
i.i.d. subsequences. 
In this section, we modify the context quantization scheme and apply the GM-based denoiser~(\ref{eq:FJ_denoiser}) to the i.i.d. subsequences, so that our universal denoiser can denoise stationary ergodic signals that are unbounded or with unknown bounds.
\subsection{Background}
\label{subsec:SW_old}
Consider the denoising problem~(\ref{eq:denoising}), where the input ${\bf x}$ is stationary ergodic. The main idea of the context quantization scheme~\cite{SW_Context2009} is
to quantize the noisy symbols ${\bf q}$ to generate quantized contexts that are used to partition the unquantized symbols into subsequences. That is, given the noisy observations ${\bf q}\in\mathbb{R}^N$, define the context of $q_j$ as
${\bf c}_j=[{\bf q}_{j-k}^{j-1};{\bf q}_{j+1}^{j+k}]\in\mathbb{R}^{2k}$ 
for $j=1+k,...,N-k$, 
where $[ {\bf a};{\bf b} ]$ denotes the concatenation 
of the sequences ${\bf a}$ and ${\bf b}$. 
For $j\le k$ or $j\ge N-k+1$, the median value $q_{\text{med}}$ of ${\bf q}$ is used as the missing symbols in the contexts. 
As an example for $j=k$, we only have $k-1$ symbols in ${\bf q}$ before $q_k$, and so the first symbol in ${\bf c}_k$ is missing; we define ${\bf c}_k=[q_{\text{med}};{\bf q}_{1}^{k-1};{\bf q}_{k+1}^{2k}]$.
Vector quantization can then be applied to the context set $\mathcal{C}=\lbrace {\bf c}_j:j=1,...,N \rbrace$, and each ${\bf c}_j$ is assigned a label $l_j\in\lbrace  1,...,L\rbrace
$ that represents the cluster that ${\bf c}_j$ belongs to. Finally, the $L$ subsequences that consist of symbols from ${\bf q}$ with the same label are obtained by taking ${\bf q}^{(l)}=\lbrace q_j:l_j=l \rbrace$, for $l=1,...,L$.

The symbols in each subsequence  ${\bf q}^{(l)}$ are regarded as approximately
conditionally identically distributed given the common quantized 
contexts.
The rationale underlying this concept is that a sliding-window denoiser uses information from the contexts to estimate the current symbol, and symbols with similar contexts in the noisy output
of the scalar channel have similar contexts in the original signal.
Therefore, symbols with similar contexts can be grouped together 
and denoised using the same denoiser.
Note that Sivaramakrishnan and Weissman~\cite{SW_Context2009} propose a second subsequencing step, which further
partitions each subsequence into smaller subsequences such
that a symbol in a subsequence does not belong to the contexts
of any other symbols in this subsequence. This step ensures
that the symbols within each subsequence are mutually
independent, which is crucial for theoretical analysis.
However, for finite-length signals, small subsequences may occur, and they
may not contain enough symbols to
learn its empirical pdf well. Therefore, we omit this second subsequencing step
in our implementations.

In order to estimate the distribution of ${\bf x}^{(l)}$, which is the clean subsequence corresponding to ${\bf q}^{(l)}$, Sivaramakrishnan and Weissman~\cite{SW_Context2009} first estimate the pdf $\widehat{p}_{Q}^{(l)}$ of ${\bf q}^{(l)}$ via kernel density estimation. They then quantize the range that {\em $x_i$}'s take values from and the levels of the empirical distribution function of ${\bf x}$, and find a quantized distribution function that matches $\widehat{p}_{Q}^{(l)}$ well.
Once the distribution function of ${\bf x}^{(l)}$ is obtained, 
the conditional expectation of the symbols in the $l$-th subsequence can be calculated.

For error metrics that satisfy some mild technical conditions, 
Sivaramakrishnan and Weissman~\cite{SW_Context2009} have proved
for stationary ergodic signals with bounded components that 
their universal denoiser asymptotically achieves the optimal estimation error among all sliding-window denoising schemes despite not knowing the prior for the signal. When the error metric is square error, the optimal error is the MMSE.

\subsection{Extension to unbounded signals and signals with unknown bounds}
\label{subsec:SW_new}
Sivaramakrishnan and Weissman~\cite{SW_Context2009} have shown that one can denoise a stationary ergodic signal by 
({\em i}) grouping together symbols with similar contexts and ({\em ii}) applying an i.i.d. denoiser to each group. 
Such a scheme is optimal in the limit of large signal dimension $N$. However, their denoiser assumes an input with known bounds, 
which might make it inapplicable to some real-world settings.
In order to be able to estimate signals that take values from the entire real line,  in step ({\em ii}), we apply the GM learning algorithm for density estimation, which has been discussed in detail in Section~\ref{sec:FJ}, and compute the conditional expectation with the estimated density as our i.i.d. denoiser.

We now provide details about a modification made to step ({\em i}).
The context set $\mathcal{C}$ is acquired in the same way as described in Section~\ref{subsec:SW_old}.
Because the symbols in the context ${\bf c}_j\in\mathcal{C}$ that are closer in index to $q_j$ are likely to provide more information about $x_j$ than the ones that are located further away,
we add weights to the contexts before clustering. 
That is, for each ${\bf c}_j\in\mathcal{C}$ of length $2k$, the weighted context is defined as 
\begin{equation*}
{\bf c}_j'={\bf c}_j\odot{\bf w}, 
\end{equation*}
where $\odot$ denotes a point-wise product, and the weights take values, 
\begin{equation}
\label{eq:weight}
w_{k_i} =
\begin{cases}
e^{-\beta (k-k_i)},\quad k_i=1,..,k\\
e^{-\beta (k_i-k-1)},\quad k_i=k+1,...,2k
\end{cases},
\end{equation}
for some $\beta\ge 0$. 
While in noisier channels, it might be necessary to use information from longer contexts, comparatively short contexts could be sufficient for cleaner channels. Therefore,
the exponential decay rate $\beta$ is made adaptive to the noise level in a way such that $\beta$ increases with SNR. Specifically, $\beta$ is chosen to be linear in SNR:
\begin{equation}
\label{eq:weight_exp}
\beta = b_1\log_{10}((\|{\bf q}\|_2^2/N-\sigma_v^2)/\sigma_v^2)+b_2,
\end{equation}
where $b_1>0$ and $b_2$ can be determined numerically. 
Specifically, we run the algorithm with a sufficiently large range of $\beta$ values for various input signals at various SNR levels and 
measurement rates. For each setting, we select the $\beta$ that achieves the best reconstruction quality. Then, $b_1$ and $b_2$ are obtained using the least squares approach. Note that $b_1$ and $b_2$ are fixed for all the simulation results presented in Section~\ref{sec:numerical}.
If the parameters were tuned for each individual input signal, then the optimal parameter values might vary for different input signals, and
the reconstruction quality might be improved. The simulation results in Section~\ref{sec:numerical} with fixed parameters show that when the
parameters are slightly off from the individually optimal ones, the reconstruction quality of AMP-UD is still comparable
or better than the prior art.
We choose the linear relation because it is simple and fits well with our empirical optimal values for $\beta$; other choices for $\beta$ might be possible.
The weighted context set $\mathcal{C}'=\lbrace {\bf c}'_j:j=1,...,N \rbrace$ is then sent to a $k$-means algorithm~\cite{MacQueen1967kmeans}, and ${\bf q}^{(l)}, l=1,...,L,$ are obtained according to the labels determined via clustering. 
We can now apply the GM-based i.i.d. denoiser~(\ref{eq:FJ_denoiser}) to each subsequence separately. However, one potential problem is that the GM fitting algorithm might not provide a good estimate of the model when the number of data points is small. We propose two approaches to address this small cluster issue.

{\bf Approach 1: Borrow members from nearby clusters.}
A post-processing step can be added to ensure that the
pdf of ${\bf q}^{(l)}$ is estimated from no less than $T$ symbols. That is, if the size of ${\bf q}^{(l)}$, which is denoted by $B$, is less than $T$, then $T-B$ symbols in other clusters whose contexts are closest to the centroid of the current cluster are included to estimate the empirical pdf of ${\bf q}^{(l)}$, while after the pdf is estimated, the extra symbols are removed, and only ${\bf q}^{(l)}$ is denoised with the currently estimated pdf.
We call UD with Approach 1 ``UD1."{\footnote{A related approach is k-nearest neighbors, where for each symbol in ${\bf q}$, we find $T$ symbols whose contexts are nearest to that of the current symbol and estimate its pdf from the T symbols. The k-nearest neighbors approach requires to run the GM learning algorithm~\cite{FigueiredoJain2002} $N$ times in each AMP iteration, which significantly slows down the algorithm. }

{\bf Approach 2: Merge statistically similar subsequences.} An alternative approach is to merge subsequences iteratively according to their statistical characterizations. 
The idea is to find subsequences with pdfs that are close in Kullback-Leibler (KL) distance~\cite{Cover06}, and decide whether merging them can yield a better model according to the  minimum description length (MDL)~\cite{BRY98} criterion.
Denote the iteration index for the merging process by $h$.
After the $k$-means algorithm, we have obtained a set of subsequences $\{{\bf q}_{h}^{(l)}: l=1,...,L_h\}$, where $L_h$ is the current number of subsequences. A GM pdf $\widehat{p}_{Q,h}^{(l)}$ is learned for each subsequence ${\bf q}_{h}^{(l)}$. 
The MDL cost $c^\text{MDL}_h$ for the current model is calculated as:
\begin{equation*}
c^\text{MDL}_h = 
-\sum_{l=1}^{L_h} \sum_{i=1}^{|{\bf q}_{h}^{(l)}|}\log \left(\widehat{p}_{Q,h}^{(l)}( q_{i,h}^{(l)})\right)
+\sum_{l=1}^{L_h} \frac{3\cdot m_{h}^{(l)}}{2}\log\left(|{\bf q}_{h}^{(l)}|\right)
+2\cdot L_h
+L_0\cdot\sum_{l=1}^{L_h} \frac{n_{h}^{(l)}}{L_0}\log\left(\frac{L_0}{n_{h}^{(l)}} \right),
\end{equation*}
where $q_{i,h}^{(l)}$ is the $i$-th entry of the subsequence ${\bf q}_h^{(l)}$, $m_{h}^{(l)}$ is the number of Gaussian components in the mixture model for subsequence ${\bf q}_{h}^{(l)}$, $L_0$ is the number of subsequences before the merging procedure, and $n_{h}^{(l)}$ is the number of subsequences in the initial set $\{{\bf q}_{0}^{(l)}: l=1,...,L_0\}$ that are merged to form the subsequence ${\bf q}_{h}^{(l)}$. 
The four terms in $c^\text{MDL}_{h}$ are interpreted as follows. 
The first term is the negative log likelihood of the entire noisy sequence ${\bf q}_h$ given the current GM models. 
The second term is the penalty for the number of parameters used to describe the model, where we have 3 parameters $(\alpha,\mu,\sigma^2)$ for each Gaussian component, and $m_{h}^{(l)}$ components for the subsequence ${\bf q}_{h}^{(l)}$. 
The third term arises from 2 bits that are used to encode $m_{h}^{(l)}$ for $l=1,...,L_h$, because our numerical results have shown that the number of Gaussian components rarely exceeds 4.
In the fourth term, $\sum_{l=1}^{L_h} \frac{n_{h}^{(l)}}{L_0}\log\left(\frac{L_0}{n_{h}^{(l)}} \right)$ is the uncertainty that a subsequence from the initial set is mapped to ${\bf q}_{h}^{(l)}$ with probability $n_{h}^{(l)}/L_0$, for $l=1,...,L_h$. Therefore, the fourth term is the coding length for mapping the $L_0$ subsequences from the initial set to the current set.

We then compute the KL distance between
the pdf of ${\bf q}_{h}^{(s)}$ and that of ${\bf q}_{h}^{(t)}$, for $s,t=1,...,L_h$: 
\begin{equation*}
D\left(\left.\widehat{p}_{Q,h}^{(s)}\right\| \widehat{p}_{Q,h}^{(t)}\right)=\int \widehat{p}_{Q,h}^{(s)}(q)\log\left(\frac{\widehat{p}_{Q,h}^{(s)}(q)}{\widehat{p}_{Q,h}^{(t)}(q)}\right)\text{d}q.
\end{equation*}
A symmetric $L_h\times L_h$ distance matrix ${\bf D}_h$ is obtained by 
letting its $s$-th row and $t$-th column be
\begin{equation*}
D\left(\left.\widehat{p}_{Q,h}^{(s)}\right\| \widehat{p}_{Q,h}^{(t)}\right)+D\left(\left.\widehat{p}_{Q,h}^{(t)}\right\| \widehat{p}_{Q,h}^{(s)}\right).
\end{equation*}
Suppose the smallest entry in the upper triangular part of ${\bf D}_h$ (not including the diagonal) is located in the $s^*$-th row and $t^*$-th column, then ${\bf q}_h^{(s^*)}$ and ${\bf q}_h^{(t^*)}$ are temporarily merged to form a new subsequence, and a new GM pdf is learned for the merged subsequence. 
We now have a new model with $L_{h+1}=L_{h}-1$ GM pdfs, and the MDL criterion $c_{h+1}^\text{MDL}$ is calculated for the new model. If  $c_{h+1}^\text{MDL}$ is smaller than $c_{h}^\text{MDL}$, then we accept the new model, and calculate a new $L_{h+1}\times L_{h+1}$ distance matrix ${\bf D}_{h+1}$;
otherwise we keep the current model, and look for the next smallest entry in the upper triangular part of the current $L_h\times L_h$ distance matrix.
The number of subsequences is decreased by at most one after each iteration, and the merging process ends when there is only one subsequence left, or the smallest KL distance between two GM pdfs is greater than some threshold, which is determined numerically. 
We call UD with Approach 2 ``UD2."

We will see in Section~\ref{sec:numerical} that UD2 is more reliable than UD1 in terms of MSE performance, whereas UD1 is 
faster
than UD2. 
This is because UD2 applies a more complicated (and thus slower) subsequencing procedure, which allows more accurate GM models to be fitted to subsequences.

\section{Proposed Universal CS recovery algorithm}
\label{sec:proposed_algo}
Combining the three components that have been discussed in 
Sections~\ref{sec:AMP}--\ref{sec:SW},
we are now ready to introduce our proposed universal CS recovery algorithm AMP-UD.
Note that the AMP-UD algorithm is designed for medium to large size problems. Specifically, the problem size should be large enough, such that the decoupling effect of AMP, which converts the compressed sensing problem to a series of scalar channel denoising problems, approximately holds, and that the statistical information about the input can be approximately estimated by the universal denoiser.

Consider a linear system~(\ref{eq:matrix_channel}), 
where the input signal ${\bf x}$ is stationary and ergodic with unknown distributions, and the matrix ${\bf A}$ has i.i.d. Gaussian entries. To estimate ${\bf x}$ from ${\bf y}$ given ${\bf A}$, 
we apply AMP as defined in~(\ref{eq:AMPiter1}) and (\ref{eq:AMPiter2}). In each iteration, 
AWGN corrupted observations,
${\bf q}_t={\bf x}_t+{\bf A}^T{\bf r}_t={\bf x}+{\bf v}$, are obtained, 
{where $\sigma_v^2$ is estimated by $\widehat{\sigma}_t^2$~(\ref{eq:AMP_noise_est}).}
A subsequencing approach is applied to generate i.i.d. subsequences, where 
Approach 1 and Approach 2~(Section~\ref{subsec:SW_new}) are two possible implementations.
The GM-based i.i.d. denoiser~(\ref{eq:FJ_denoiser}) is then utilized to 
denoise each i.i.d. subsequence.

To obtain the Onsager correction term in (\ref{eq:AMPiter2}), we need to calculate the derivative of $\eta_{\text{iid}}$ (\ref{eq:FJ_denoiser}). For $q\in\mathbb{R}$, denoting
\begin{align*}
f(q) &= \sum_{s=1}^S\alpha_s\mathcal{N}(q;\mu_s,\sigma_s^2+\sigma_v^2)\left(\frac{\sigma_s^2}{\sigma_s^2+\sigma_v^2}(q-\mu_s)+\mu_s\right),\\
g(q) &= \sum_{s=1}^S\alpha_s\mathcal{N}(q;\mu_s,\sigma_s^2+\sigma_v^2),
\end{align*}
we have that
\begin{align*}
f'(q) &= \sum_{s=1}^S\alpha_s\mathcal{N}(q;\mu_s,\sigma_s^2+\sigma_v^2)\cdot\left(\frac{\sigma_s^2+\mu_s^2-q\mu_s}{\sigma_s^2+\sigma_v^2}-\left(\frac{\sigma_s(q-\mu_s)}{\sigma_s^2+\sigma_v^2}\right)^2\right),\\
g'(q) &= \sum_{s=1}^S\alpha_s\mathcal{N}(q;\mu_s,\sigma_s^2+\sigma_v^2)\left(- \frac{q-\mu_s}{\sigma_s^2+\sigma_v^2}\right).
\end{align*}
Therefore,
\begin{equation}
\label{eq:denoiser_FJ_deriv}
\eta_{\text{iid}}'(q)=\frac{f'(q)g(q)-f(q)g'(q)}{(g(q))^2}.
\end{equation}

We highlight that AMP-UD is unaware of the input SNR and also unaware of the input statistics. The noise variance $\sigma_v^2$ in the
scalar channel denoising problem is estimated by the average energy of the residual~(\ref{eq:AMP_noise_est}). The input's statistical structure is learned by the universal denoiser without any prior assumptions. 

It has been proved~\cite{SW_Context2009} that the context quantization universal denoising scheme can asymptotically achieve the MMSE for stationary ergodic signals with known bounds. 
We have extended the scheme to unbounded signals in Sections~\ref{sec:FJ}~and~\ref{sec:SW}, and conjecture that our modified universal denoiser can asymptotically achieve the MMSE for unbounded stationary ergodic signals.
AMP with MMSE-achieving separable denoisers has been proved to asymptotically achieve the MMSE in linear systems for i.i.d. inputs~\cite{Bayati2011}. In Section~\ref{subsec:SE}, we have provided numerical evidence that shows that SE holds for AMP with Bayesian sliding-window denoisers. 
Bayesian sliding-window denoisers with proper window-sizes are MMSE-achieving non-separable denoisers~\cite{SW_Context2009}. 
Given that our universal denoiser resembles a Bayesian sliding-window denoiser, we conjecture that AMP-UD can achieve the MMSE for stationary ergodic inputs in the limit of large linear systems where the matrix has i.i.d. random entries.
Note that we have optimized the window-size for inputs of length $N=10000$ via numerical experiments. We believe that the window size should increase with $N$, and leave the characterization of the optimal window size for future work.

\begin{figure}[t!]
\center
\includegraphics[width=85mm]{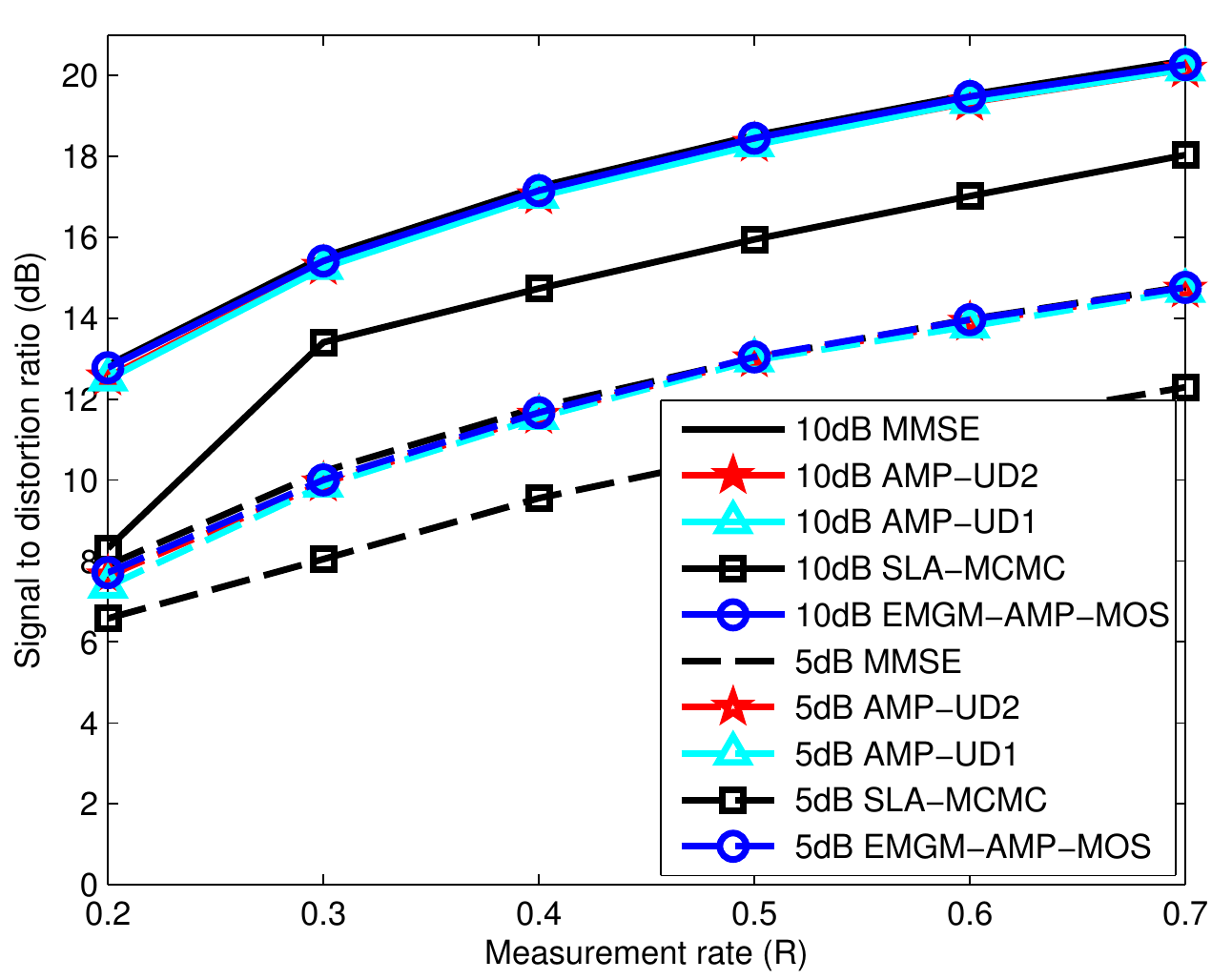}
\caption{Two AMP-UD implementations, SLA-MCMC, and EM-GM-AMP-MOS reconstruction results for an i.i.d. sparse Laplace signal as a function of measurement rate ($R=M/N$). Note that the SDR curves for the two AMP-UD implementations and EM-GM-AMP-MOS overlap the MMSE curve. ($N=10000$, SNR = 5 dB or 10 dB.)}
\label{fig:AMP_Laplace}
\end{figure}

\section{Numerical results}
\label{sec:numerical}
We run AMP-UD1 (AMP with UD1) and AMP-UD2 (AMP with UD2) in MATLAB
on a Dell OPTIPLEX 9010 running an Intel(R) 
$\text{Core}^{\text{TM}}$ i7-3770 with 16GB RAM, and test them utilizing different types of signals, including synthetic signals, a chirp sound clip, and a speech signal, at various measurement rates and SNR levels, where we remind the reader that SNR is defined in Section~\ref{subsec:SE}.  The input signal length $N$ is 10000 for synthetic signals and roughly 10000 for the chirp sound clip and the speech signal.
The context size $2k$ is chosen to be 12, and the contexts are weighted according to (\ref{eq:weight}) and (\ref{eq:weight_exp}). 
The context quantization is implemented via the $k$-means algorithm~\cite{MacQueen1967kmeans}.
In order to avoid possible divergence of AMP-UD, possibly due to a bad GM fitting, we employ a damping technique~\cite{Rangan2014ISIT} to slow down the evolution. Specifically, damping is an extra step in the AMP iteration~(\ref{eq:AMPiter2}); instead of updating the value of ${\bf x}_{t+1}$ by the output of the denoiser $\eta_t({\bf A}^T{\bf r}_t+{\bf x}_t)$, a weighted sum of $\eta_t({\bf A}^T{\bf r}_t+{\bf x}_t)$ and ${\bf x}_t$ is taken as follows,
\begin{equation*}
{\bf x}_{t+1}=\lambda\eta_t({\bf A}^T{\bf r}_t+{\bf x}_t)+(1-\lambda){\bf x}_t,
\end{equation*} 
for some $\lambda\in(0,1]$.
 
{\bf Parameters for AMP-UD1:}
The number of clusters $L$ is initialized as 10, and may become smaller if empty clusters occur. The lower bound $T$ on the number of symbols required to learn the GM parameters is $256$.
The damping parameter $\lambda$ is 0.1, and we run 100 AMP iterations.

{\bf Parameters for AMP-UD2:}
The initial number of clusters is set to be 30, and these clusters will be merged according to the scheme described in Section~\ref{sec:SW}. 
Because each time when merging occurs, we need to apply the GM fitting algorithm one more time to learn a new mixture model for the merged cluster, which is computationally demanding, we apply adaptive damping~\cite{Vila2015} to reduce the number of iterations required; the number of AMP iterations is set to be 30. The damping parameter is initialized to be 0.5, and will increase (decrease) within the range $[0.01,0.5]$ if the value of the scalar channel noise estimator $\widehat{\sigma}_t^2$~(\ref{eq:AMP_noise_est}) decreases (increases).

\begin{figure}[t!]
\center
\includegraphics[width=85mm]{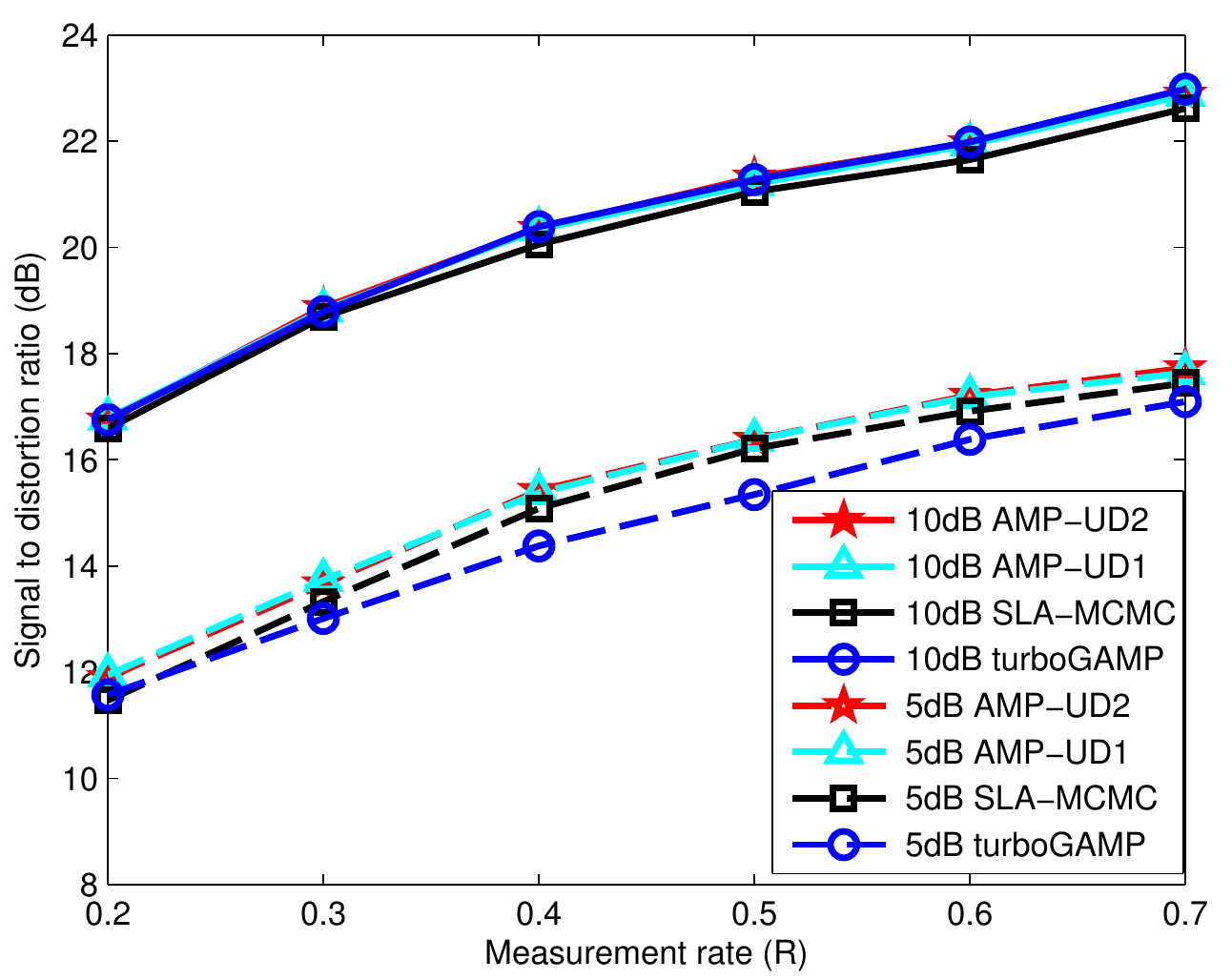}
\caption{Two AMP-UD implementations, SLA-MCMC, and turboGAMP reconstruction results for a two-state Markov signal with nonzero entries drawn from a uniform distribution $U[0,1]$ as a function of measurement rate. Note that the SDR curves for the two AMP-UD implementations overlap at SNR = 5 dB, and they both overlap turboGAMP at SNR = 10 dB.
($N=10000$, SNR = 5 dB or 10 dB.)}
\label{fig:AMP_MUnif}
\end{figure}

The recovery performance is evaluated by signal to distortion ratio ($\text{SDR}=10\log_{10}(\mathbb{E}[X^2]/\text{MSE})$), where the MSE is averaged over 50 random draws of ${\bf x}$, ${\bf A}$, and ${\bf z}$.

We compare the performance of the two AMP-UD implementations to 
({\em i}) 
the universal CS recovery algorithm SLA-MCMC~\cite{ZhuBaronDuarte2014_SLAM}; and ({\em ii}) the empirical Bayesian message passing approaches EM-GM-AMP-MOS~\cite{EMGMTSP} for i.i.d. inputs and turboGAMP~\cite{turboGAMP} for non-i.i.d. inputs. 
Note that EM-GM-AMP-MOS 
assumes during recovery that the input is i.i.d., 
whereas turboGAMP is designed for non-i.i.d. inputs 
with a known statistical model.
We do not include results for other well-known CS
algorithms such as compressive sensing matching pursuit 
(CoSaMP)~\cite{Cosamp08}, gradient projection for sparse 
reconstruction (GPSR)~\cite{GPSR2007}, or $\ell_1$ minimization~\cite{DonohoCS,CandesRUP}, because their SDR
performance is consistently weaker than the three algorithms
being compared.

{\bf Sparse Laplace signal (i.i.d.):}
We tested i.i.d. sparse Laplace signals that follow the distribution $p_X(x)=0.03\mathcal{L}(0,1)+0.97\delta(x)$, where $\mathcal{L}(0,1)$ denotes a Laplacian distribution with mean zero and variance one, and $\delta(\cdot)$ is the delta function~\cite{Papoulis91}. 
It is shown in Fig.~\ref{fig:AMP_Laplace} that the two AMP-UD implementations and EM-GM-AMP-MOS achieve the MMSE~\cite{GuoBaronShamai2009,RFG2012}, whereas SLA-MCMC has 
weaker performance, because the MCMC approach is expected to sample 
from the posterior and its MSE is twice the 
MMSE~\cite{DonohoKolmogorov,ZhuBaronDuarte2014_SLAM}.

{\bf Markov-uniform signal:}
Consider the two-state Markov state machine defined in Section~\ref{subsec:SE} with $p_{01}=\frac{3}{970}$ and $p_{10}=\frac{1}{10}$. A Markov-uniform signal (MUnif for short) follows a uniform distribution $U[0,1]$ at 
the nonzero state $s_1$. 
These parameters lead to 3\% nonzero entries in an MUnif signal on average.
It is shown in Fig.~\ref{fig:AMP_MUnif} that at low SNR, 
the two AMP-UD implementations achieve higher SDR than SLA-MCMC and turboGAMP.  
At high SNR, the two AMP-UD implementations and turboGAMP have similar SDR performance, and are slightly better than SLA-MCMC. We highlight that turboGAMP
needs side information about the Markovian structure of the signal, whereas the two AMP-UD implementations and SLA-MCMC do not.

\begin{figure}[t!]
\center
\includegraphics[width=85mm]{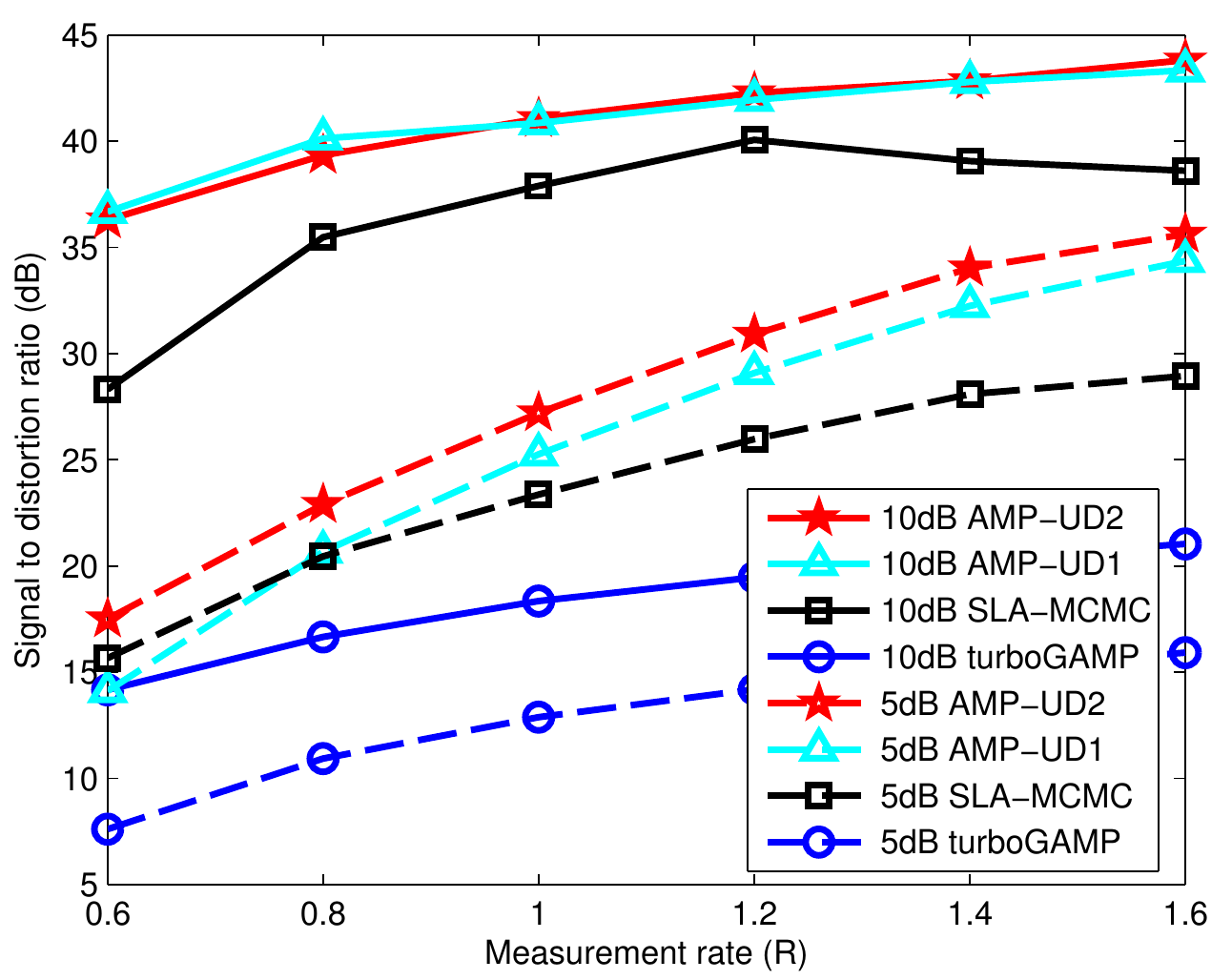}
\caption{Two AMP-UD implementations, SLA-MCMC, and turboGAMP reconstruction results for a dense two-state Markov signal with nonzero entries drawn from a Rademacher ($\pm 1$) distribution as a function of measurement rate. 
($N=10000$, SNR = 10 dB or 15 dB.)}
\label{fig:AMP_MRad}
\end{figure}

{\bf Dense Markov-Rademacher signal:}
Consider the two-state Markov state machine defined in 
Section~\ref{subsec:SE} with $p_{01}=\frac{3}{70}$ and $p_{10}=\frac{1}{10}$. 
A dense Markov Rademacher signal (MRad for short) 
takes values from $\lbrace -1,+1  \rbrace$ with equal probability 
at $s_1$. These parameters lead to 30\% nonzero entries in an 
MRad signal on average. Because the MRad signal is dense 
(non-sparse), we must measure it with somewhat larger 
measurement rates and SNRs than before.
It is shown in Fig.~\ref{fig:AMP_MRad} that 
the two AMP-UD implementations and SLA-MCMC have better overall performance than turboGAMP. 
AMP-UD1 outperforms SLA-MCMC except for the lowest tested measurement rate at low SNR,
whereas AMP-UD2 outperforms SLA-MCMC consistently.

\begin{figure}[t!]
\center
\includegraphics[width=85mm]{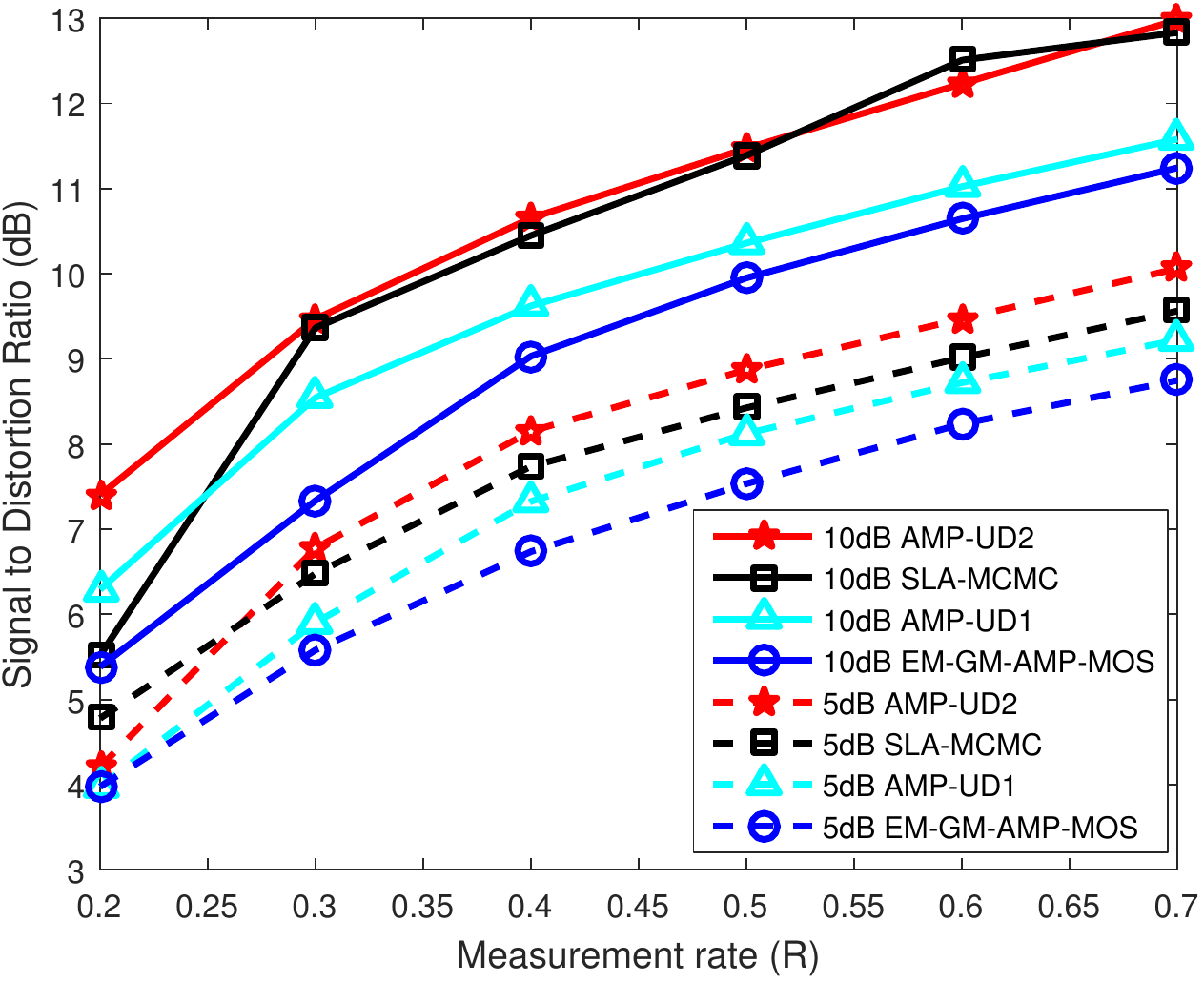}
\caption{Two AMP-UD implementations, SLA-MCMC, and EM-GM-AMP-MOS reconstruction results for a chirp sound clip as a function of measurement rate. 
($N=9600$, SNR = 5 dB or 10 dB.)}
\label{fig:AMP_Chirp}
\end{figure}

{\bf Chirp sound clip and speech signal:}
Our experiments up to this point use synthetic signals. We now evaluate the reconstruction quality of AMP-UD for two real-world signals. A ``Chirp'' sound clip and a speech signal are used. We cut a segment with length 9600 out of the ``Chirp'' and a segment with length 10560 out of the speech signal (denoted by ${\bf x}$), and performed a short-time discrete cosine transform (DCT) with window size, number of DCT points, and hop size all being 32.
The resulting short-time DCT coefficients matrix are then vectorized to form a coefficient vector ${\boldsymbol\theta}$. Denoting the short-time DCT matrix by ${\bf W}^{-1}$, we have ${\boldsymbol\theta}={\bf W}^{-1}{\bf x}$. Therefore, we can rewrite~(\ref{eq:matrix_channel}) as ${\bf y}={\bf \Phi}{\boldsymbol\theta} +{\bf z}$,
where ${\bf \Phi}={\bf AW}$. Our goal is to reconstruct ${\boldsymbol\theta}$ from the measurements ${\bf y}$ and the matrix ${\bf \Phi}$. After we obtain the estimated coefficient vector $\widehat{\boldsymbol\theta}$, the estimated signal is calculated as $\widehat{{\bf x}}={\bf W}\widehat{\boldsymbol\theta}$.
Although the coefficient vector ${\boldsymbol\theta}$ may exhibit some type of memory, it is not readily modeled in closed form, and so we cannot provide a valid model for turboGAMP~\cite{turboGAMP}. Therefore, we use EM-GM-AMP-MOS~\cite{EMGMTSP} instead of turboGAMP~\cite{turboGAMP}. 
The SDRs for the two AMP-UD implementations, SLA-MCMC and EM-GM-AMP-MOS~\cite{EMGMTSP} for the ``Chirp" are plotted in Fig.~\ref{fig:AMP_Chirp} and the speech signal in Fig.~\ref{fig:AMP_speech}.
We can see that both AMP-UD implementations outperform EM-GM-AMP-MOS consistently, which implies that the simple i.i.d. model is suboptimal for these two real-world signals. 
Moreover, AMP-UD2 provides comparable and in most cases higher SDR than SLA-MCMC, which indicates that AMP-UD2 is more reliable in learning various statistical structures than SLA-MCMC.
AMP-UD1 is the fastest among the four algorithms, but it may have lower reconstruction quality than AMP-UD2 and SLA-MCMC, owing to poor selection of the subsequences.
It is worth mentioning that we have also run simulations on an electrocardiograph (ECG) signal, and EM-GM-AMP-MOS achieved similar SDR as the two AMP-UD implementations, which indicates that an i.i.d. model might be adequate to represent the coefficients of the ECG signal; the plot is omitted for brevity.

{\bf Runtime:} The runtime of AMP-UD1 and AMP-UD2 for MUnif, MRad, and the speech signal is typically under 5 minutes and 10 minutes, respectively, but somewhat more for signals such as sparse Laplace and the chirp sound clip that require a large number of Gaussian components to be fit. For comparison, the runtime of SLA-MCMC is typically an hour, whereas typical runtimes of EM-GM-AMP-MOS and turboGAMP are 30 minutes. To further accelerate AMP, we could consider parallel computing. That is, after clustering, the Gaussian mixture learning algorithm can be implemented simultaneously in different processors.

\section{Conclusion}
\label{sec:conclusion}
In this paper, we introduced a universal 
compressed sensing recovery algorithm AMP-UD that applies our proposed
universal denoiser (UD) within approximate message passing (AMP).
AMP-UD is designed to reconstruct stationary ergodic signals from
noisy linear measurements. 
The performance of two AMP-UD implementations was evaluated via simulations, where it was shown that AMP-UD
achieves favorable signal to distortion ratios compared to existing universal algorithms, 
and that its runtime is typically faster.

\begin{figure}[t!]
\center
\includegraphics[width=85mm]{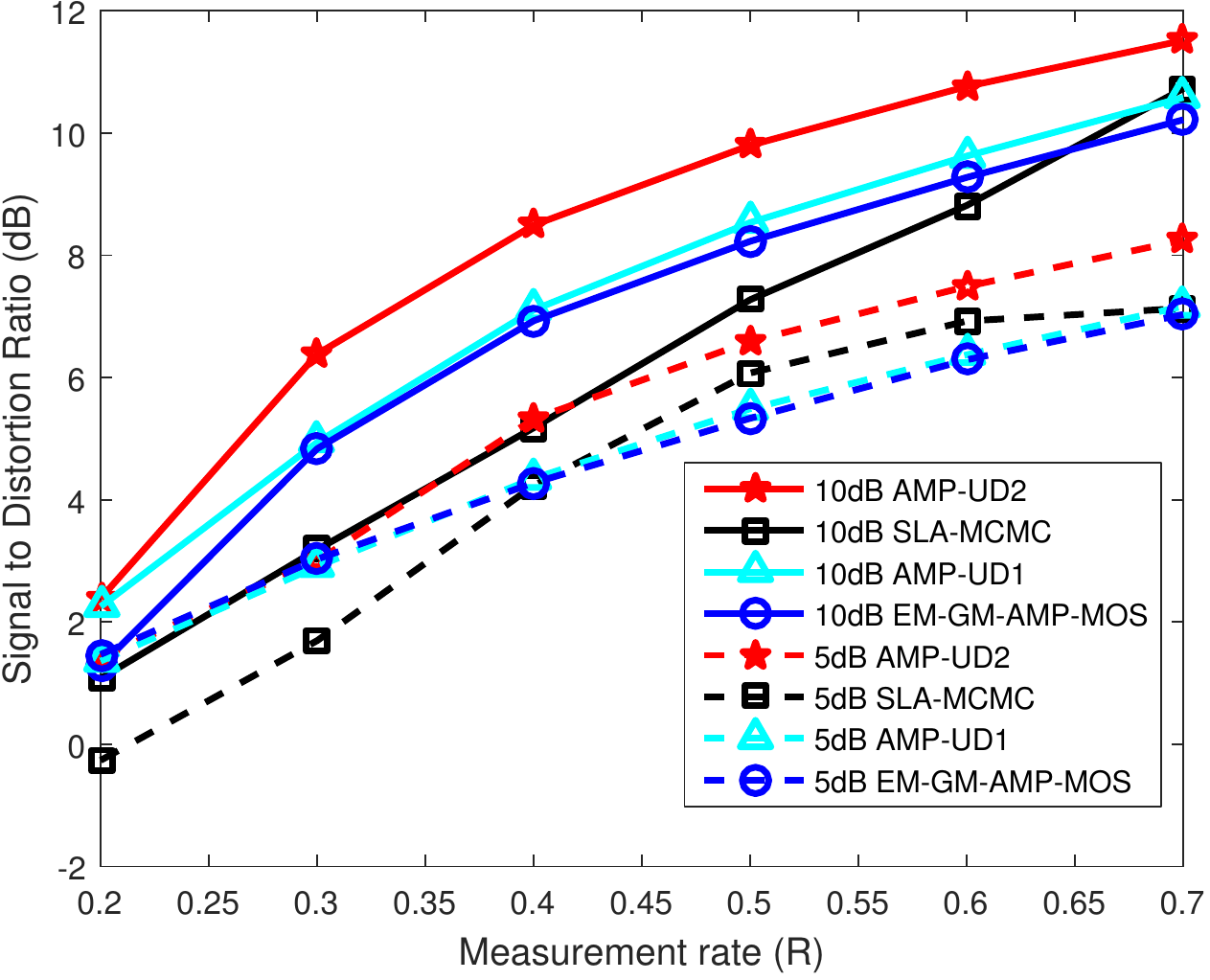}
\caption{Two AMP-UD implementations, SLA-MCMC, and EM-GM-AMP-MOS reconstruction results for a speech signal as a function of measurement rate. 
($N=10560$, SNR = 5 dB or 10 dB.)}
\label{fig:AMP_speech}
\end{figure}

AMP-UD combines three existing schemes:
({\em i}) AMP~\cite{DMM2009}; 
({\em ii}) universal denoising~\cite{SW_Context2009}; and
({\em iii}) a density estimation approach based on Gaussian mixture (GM) fitting~\cite{FigueiredoJain2002}.
In addition to the algorithmic framework, we provided three specific contributions. 
First, we provided numerical results showing that SE holds for non-separable Bayesian sliding-window denoisers.
Second, we modified the GM learning algorithm, and extended it to an i.i.d. denoiser.
Third, we designed a universal denoiser that does not need to know the bounds of the input or require the input signal to be bounded. Two implementations of the universal denoiser were provided, with one being faster and the other achieving better reconstruction quality in terms of signal to distortion ratio.

There are numerous directions for future work.
First, our current algorithm was designed to minimize the square error, and the denoiser could be modified to minimize
other error metrics~\cite{Tan2014}.
Second, AMP-UD was designed to reconstruct one-dimensional signals. In order to support applications that process multi-dimensional signals such as images, it might be instructive to employ universal image denoisers within AMP.
Third, the relation between the input length and the optimal window-size, as well as the exponential decay rate of the context weights, can be investigated.
Finally, we can modify our work to support measurement noise with unknown distributions as an extension to adaptive generalized AMP~\cite{Kamilov2014}.

\section*{Appendix}
We follow the derivation in Figueiredo and Jain~\cite{FigueiredoJain2002}. Denoting $\boldsymbol\theta=\{\alpha_s,\mu_s,\sigma_s^2\}_{s=1}^{S}$, the MML-based criterion is 
\begin{equation}
\mathcal{L}({\bf q},\boldsymbol\theta)=\frac{n}{2}\sum\limits_{s:\alpha_s>0}\log(N\alpha_s)
+\frac{S_{nz}}{2}\log(N)-\log\left(p({\bf q}|\boldsymbol\theta)\right),\label{eq:MML_cost}
\end{equation}
where $n=2$ is the number of parameters per Gaussian component, and $S_{nz}$ is the number of components with nonzero mixing probability $\alpha_s$. 
The first term is the coding length of $\{\mu_s,\sigma_s^2\}_{s=1}^{S_{nz}}$, because the expected number of data points that are from the $s$-th component is $N\alpha_s$, hence the effective sample size for estimating $\{\mu_s,\sigma_s^2\}$ is $N\alpha_s$. The second term is the coding length of $\alpha_s$'s, because $\alpha_s$'s are estimated from $N$ data points. The third term is the coding length of the data sequence ${\bf q}$.

The {\em complete data} expression for $\log\left(p({\bf q}|\boldsymbol\theta)\right)$ is
\begin{align*}
\log\left(p({\bf q},{\bf x},{\bf z}|\boldsymbol\theta)\right)
&=\sum_{i=1}^N \log\left(p(q_i,x_i,z_i^{(1)},...,z_i^{(S)}|\boldsymbol\theta)\right)\\
&=\sum_{i=1}^N \log\left(p(q_i,x_i|z_i^{(1)},...,z_i^{(S)},\boldsymbol\theta)p(z_i^{(1)},...,z_i^{(S)})\right)\\
&=\sum_{i=1}^N \log\left(\prod_{s=1}^k p(q_i,x_i|\{\mu_s,\sigma_s^2\})^{z_i^{(s)}}\prod_{s=1}^S \alpha_s^{z_i^{(s)}}\right)\\
&=\sum_{i=1}^N \log\left(\prod_{s=1}^S \left(\alpha_s p(q_i,x_i|\{\mu_s,\sigma_s^2\}\right))^{z_i^{(s)}}\right)\\
&=\sum_{i=1}^N\sum_{s=1}^S z_i^{(s)} \log\left(\alpha_s p(q_i,x_i|\{\mu_s,\sigma_s^2\})\right)\\
&=\sum_{i=1}^N\sum_{s=1}^S z_i^{(s)} \log\left(\alpha_s p(q_i|x_i)p(x_i|\{\mu_s,\sigma_s^2\})\right)\\
&=\sum_{i=1}^N\sum_{s=1}^S z_i^{(s)} \log\left(\alpha_s \mathcal{N}(q_i;x_i,\sigma_v^2)\mathcal{N}(x_i;\mu_s,\sigma_s^2)\right).
\end{align*} 
Replace $\log\left(p({\bf q}|\boldsymbol\theta)\right)$ in (\ref{eq:MML_cost}) with $\log\left(p({\bf q},{\bf x},{\bf z}|\boldsymbol\theta)\right)$:
\begin{equation}
\mathcal{L}({\bf q,x,z},\boldsymbol\theta)=\frac{n}{2}\sum\limits_{s:\alpha_s>0}\log(N\alpha_s)
+\frac{S_{nz}}{2}\log(N)-\sum_{i=1}^N\sum_{s=1}^S z_i^{(s)} \log\left(\alpha_s \mathcal{N}(q_i;x_i,\sigma_v^2)\mathcal{N}(x_i;\mu_s,\sigma_s^2)\right).
\end{equation}

Suppose $\widehat{\boldsymbol\theta}(t)=(\widehat{\alpha}_1(t),...,\widehat{\alpha}_S(t),\widehat{\mu}_1(t),...,\widehat{\mu}_S(t),\widehat{\sigma}_1^2(t),...,\widehat{\sigma}_S^2(t))$ is the estimate of the $\boldsymbol\theta$ at the $t$-th iteration.
\begin{align*}
&\mathbb{E}[\log\left(p({\bf q},{\bf X},{\bf Z}|\boldsymbol\theta)\right)|{\bf q},\widehat{\boldsymbol\theta}(t)]\\
&=\sum_{i=1}^N\sum_{s=1}^S \log\left(\alpha_s\right)\mathbb{E}\left[Z_i^{(s)}|{\bf q},\widehat{\boldsymbol\theta}(t)\right]\\
&+\sum_{i=1}^N\sum_{s=1}^S \mathbb{E}\left[Z_i^{(s)}\left(-\frac{1}{2}\log(2\pi\sigma_v^2)-\frac{(X_i-q_i)^2}{2\sigma_v^2}\right)|{\bf q},\widehat{\boldsymbol\theta}(t)\right]\\
&+\sum_{i=1}^N\sum_{s=1}^S \mathbb{E}\left[Z_i^{(s)}\left(-\frac{1}{2}\log(2\pi\sigma_s^2)-\frac{(X_i-\mu_s)^2}{2\sigma_s^2}\right)|{\bf q},\widehat{\boldsymbol\theta}(t)\right]\\
&=\sum_{i=1}^N\sum_{s=1}^S \log\left(\alpha_s\right)\mathbb{E}\left[Z_i^{(s)}|{\bf q},\widehat{\boldsymbol\theta}(t)\right]\\
&-\frac{1}{2}\sum_{i=1}^N\sum_{s=1}^S \log(2\pi\sigma_v^2)\mathbb{E}\left[Z_i^{(s)}|{\bf q},\widehat{\boldsymbol\theta}(t)\right]-\frac{1}{2\sigma_v^2}\sum_{i=1}^n\sum_{s=1}^S \mathbb{E}\left[Z_i^{(s)}(X_i-q_i)^2|{\bf q},\widehat{\boldsymbol\theta}(t)\right]\\
&-\frac{1}{2}\sum_{i=1}^N\sum_{s=1}^S \log(2\pi\sigma_s^2)\mathbb{E}\left[Z_i^{(s)}|{\bf q},\widehat{\boldsymbol\theta}(t)\right]-\frac{1}{2}\sum_{i=1}^N\sum_{s=1}^S \frac{1}{\sigma_s^2}\mathbb{E}\left[Z_i^{(s)}(X_i-\mu_s)^2|{\bf q},\widehat{\boldsymbol\theta}(t)\right]\\
&=\sum_{i=1}^N\sum_{s=1}^S \log\left(\alpha_s\right)\mathbb{E}\left[Z_i^{(s)}|{\bf q},\widehat{\boldsymbol\theta}(t)\right]\\
&-\frac{1}{2}\sum_{i=1}^N\sum_{s=1}^S \log(2\pi\sigma_s^2)\mathbb{E}\left[Z_i^{(s)}|{\bf q},\widehat{\boldsymbol\theta}(t)\right]-\frac{1}{2}\sum_{i=1}^N\sum_{s=1}^S \frac{1}{\sigma_s^2}\mathbb{E}\left[Z_i^{(s)}(X_i-\mu_s)^2|{\bf q},\widehat{\boldsymbol\theta}(t)\right]+C,
\end{align*}
where $C$ is a constant that does not depend on $\boldsymbol\theta$.
\begin{align*}
\mathbb{E}\left[Z_i^{(s)}|{\bf q},\widehat{\boldsymbol\theta}(t)\right]&=P\left(Z_i^{(s)}=1|{\bf q},\widehat{\boldsymbol\theta}(t)\right)\\
&=\frac{\widehat{\alpha}_s(t)\mathcal{N}(q_i;\widehat{\mu}_s(t),\sigma_v^2+\widehat{\sigma}_s(t)^2)}{\sum_{m=1}^S\widehat{\alpha}_m(t)\mathcal{N}(q_i;\widehat{\mu}_m(t),\sigma_v^2+\widehat{\sigma}_m(t)^2)},\\
\mathbb{E}\left[Z_i^{(s)}X_i|{\bf q},\widehat{\boldsymbol\theta}(t)\right]&=\mathbb{E}\left[X_i|Z_i^{(s)}=1,{\bf q},\widehat{\boldsymbol\theta}(t)\right]P\left(Z_i^{(s)}=1|{\bf q},\widehat{\boldsymbol\theta}(t)\right)\\
&=\left(\frac{\widehat{\sigma}_s^2(t)}{\widehat{\sigma}_s^2(t)+\sigma_v^2}(q_i-\widehat{\mu}_s(t))+\widehat{\mu}_s(t)\right)P\left(Z_i^{(s)}=1|{\bf q},\widehat{\boldsymbol\theta}(t)\right),\\
\mathbb{E}\left[Z_i^{(s)}X_i^2|{\bf q},\widehat{\boldsymbol\theta}(t)\right]&=\mathbb{E}\left[X_i^2|Z_i^{(s)}=1,{\bf q},\widehat{\boldsymbol\theta}(t)\right]P\left(Z_i^{(s)}=1|{\bf q},\widehat{\boldsymbol\theta}(t)\right)\\
&=\left(\frac{\sigma_v^2\widehat{\sigma}_s^2(t)}{\widehat{\sigma}_s^2(t)+\sigma_v^2}+\left(\frac{\widehat{\sigma}_s^2(t)}{\widehat{\sigma}_s^2(t)+\sigma_v^2}(q_i-\widehat{\mu}_s(t))+\widehat{\mu}_s(t)\right)^2\right)P\left(Z_i^{(s)}=1|{\bf q},\widehat{\boldsymbol\theta}(t)\right).\\
\end{align*}
Denote 
\begin{align*}
w_i^{(s)}(t)&=\mathbb{E}\left[Z_i^{(s)}|{\bf q},\widehat{\boldsymbol\theta}(t)\right],\\
a_i^{(s)}(t)&=\frac{\widehat{\sigma}_s^2(t)}{\widehat{\sigma}_s^2(t)+\sigma_v^2}(q_i-\widehat{\mu}_s(t))+\widehat{\mu}_s(t),\\
v_i^{(s)}(t)&=\frac{\sigma_v^2\widehat{\sigma}_s^2(t)}{\widehat{\sigma}_s^2(t)+\sigma_v^2}.
\end{align*}
\begin{align*}
&\mathbb{E}[\log\left(p({\bf q},{\bf X},{\bf Z}|\boldsymbol\theta)\right)|{\bf q},\widehat{\boldsymbol\theta}(t)]\\
&=\sum_{i=1}^N\sum_{s=1}^S \log\left(\alpha_s\right)w_i^{(s)}(t)-\frac{1}{2}\sum_{i=1}^N\sum_{s=1}^S \log(2\pi\sigma_s^2)w_i^{(s)}(t)
-\frac{1}{2}\sum_{i=1}^N\sum_{s=1}^S \frac{w_i^{(s)}(t)}{\sigma_s^2}\left(v_i^{(s)}(t)+\left(a_i^{(s)}(t)-\widehat{\mu}_s(t)\right)^2\right)+C.
\end{align*}
Therefore,
\begin{align}
&\mathbb{E}[\mathcal{L}({\bf q,X,Z},\boldsymbol\theta)|{\bf q},\widehat{\boldsymbol\theta}(t)]\nonumber\\
&=\frac{n}{2}\sum\limits_{s:\alpha_s>0}\log(\alpha_s)-\sum_{i=1}^N\sum\limits_{s:\alpha_s>0} \log\left(\alpha_s\right)w_i^{(s)}\nonumber\\
&+\frac{1}{2}\sum_{i=1}^N\sum\limits_{s:\alpha_s>0} \log(2\pi\sigma_s^2)w_i^{(s)}(t)
+\frac{1}{2}\sum_{i=1}^N\sum\limits_{s:\alpha_s>0} \frac{w_i^{(s)}(t)}{\sigma_s^2}\left(v_i^{(s)}(t)+\left(a_i^{(s)}(t)-\widehat{\mu}_s(t)\right)^2\right)+C',\label{eq:EM_E}
\end{align}
where $C'$ is a constant that does not depend on $\boldsymbol\theta$.

Denote $Q(\boldsymbol\theta,\widehat{\boldsymbol\theta}(t))=\mathbb{E}[\mathcal{L}({\bf q,x,z},\boldsymbol\theta)|{\bf q},\widehat{\boldsymbol\theta}(t)]$.
\begin{align*}
\frac{\partial}{\partial \mu_s}Q(\boldsymbol\theta,\widehat{\boldsymbol\theta}(t))
&=-\frac{1}{2}\sum_{i=1}^N \frac{w_i^{(s)}(t)}{\sigma_s^2}\left(2\mu_s-2a_i^{(s)}(t)\right)=0\\
\widehat{\mu}_s(t+1)&=\frac{\sum_{i=1}^N w_i^{(s)}(t)a_i^{(s)}(t)}{\sum_{i=1}^Nw_i^{(s)}(t)}.\\
\frac{\partial}{\partial \sigma_s^2}Q(\boldsymbol\theta,\widehat{\boldsymbol\theta}(t))
&=-\frac{1}{2}\sum_{i=1}^N \frac{w_i^{(s)}(t)}{\sigma_s^2}
+\frac{1}{2}\sum_{i=1}^N \frac{w_i^{(s)}(t)}{\sigma_s^4}\left(v_i^{(s)}(t)+\left(a_i^{(s)}(t)-\widehat{\mu}_s(t+1)\right)^2\right)=0\\
\widehat{\sigma}_s^2(t+1)&=\frac{\sum_{i=1}^N w_i^{(s)}(t)\left(v_i^{(s)}(t)+\left(a_i^{(s)}(t)-\widehat{\mu}_s(t+1)\right)^2\right)}{\sum_{i=1}^Nw_i^{(s)}(t)}.
\end{align*}

To estimate $\{\alpha_s\}$, notice that we have the constraints $0\leq\alpha_s\leq 1, \forall s$ and $\sum_{s=1}\alpha_s=1$.
Collecting the terms in (\ref{eq:EM_E}) that contain $\{\alpha_s\}$, we have
\begin{equation*}
\sum\limits_{s:\alpha_s>0}\log(\alpha_s)\left(\frac{n}{2}-\sum_{i=1}^Nw_i^{(s)}\right)=-\log\left(\prod\limits_{s:\alpha_s>0}\alpha_s^{\sum\limits_{i=1}^N w_i^{(s)}-\frac{n}{2}}\right),
\end{equation*}
which is the negative log likelihood of a quantity that is proportional to a Dirichlet pdf of $(\alpha_s,...,\alpha_{S_{nz}})$, and its mode appears at
\begin{equation*}
\alpha_s = \frac{\sum\limits_{i=1}^N w_i^{(s)}-\frac{n}{2}}{\sum\limits_{s=1}^{S_{nz}}\left(\sum\limits_{i=1}^N w_i^{(s)}-\frac{n}{2}\right)},\quad\sum\limits_{i=1}^N w_i^{(s)}-\frac{n}{2}>0.
\end{equation*}
Hence,
\begin{equation*}
\widehat{\alpha}_s(t+1) = \frac{\max\Big\lbrace\sum\limits_{i=1}^N w_i^{(s)}(t)-\frac{n}{2},0\Big\rbrace}{\sum\limits_{s=1}^{S_{nz}}\max\Big\lbrace\sum\limits_{i=1}^N w_i^{(s)}(t)-\frac{n}{2},0\Big\rbrace}.
\end{equation*}

\section*{Acknowledgements}
We thank Mario Figueiredo and Tsachy Weissman for informative discussions; and Ahmad Beirami and Marco Duarte for detailed suggestions on the manuscript.

\ifCLASSOPTIONcaptionsoff
\newpage
\fi
\bibliographystyle{IEEEtran}
\bibliography{IEEEabrv,cites}

\end{document}